\documentclass[%
  preprint,
amsmath,amssymb,
aps,
]{revtex4-1}
\usepackage{float}
\usepackage{graphicx}
\usepackage{dcolumn}
\usepackage{bm}
 
\begin{document}
 
 
\title{A transformation optics approach to singular metasurfaces}
 
\author{Fan Yang}
\author{Paloma A. Huidobro}%
\author{John B. Pendry}
\affiliation{%
  The Blackett Laboratory \\ Department of Physics\\ Imperial College London\\ London SW7 2AZ, UK
}%
 
 
 
 
 
\begin{abstract}
Surface plasmons dominate the optical response of metal surfaces, and their nature is controlled by surface geometry. Here we study metasurfaces containing singularities in the form of sharp edges and characterized by three quantum numbers despite the two-dimensional nature of the surface. We explore the nature of the plasmonic excitations, their ability to generate large concentrations of optical energy, and the transition from the discrete excitation spectrum of a non-singular surface to the continuous spectrum of a singular metasurface.
\end{abstract}
 
\maketitle
 
 
\section{\label{sec:level1}Introduction}
 
In a previous paper \cite{pendry2017compacted} we commented on the curious mathematical structure of the spectra of singular surfaces which are characterized by three quantum numbers despite the two-dimensional nature of the surface. Transformation optics \cite{Ward1996,Pendry2006} shows how the third dimension is hidden within the singularity \cite{pendry2017compacted}. This paper will explore in detail the nature of the plasmonic excitations, their ability to generate large concentrations of optical energy, and the transition from the discrete excitation spectrum of a non-singular surface to the continuous spectrum of a singular metasurface. Sub-wavelength metal gratings couple external light into surface plasmons, efficiently localizing the electromagnetic energy and finding applications in optical bio-sensing and photovoltaics \cite{Wen2014,Munday2011,Dhawan2011,Bog2012,Kuo2012}. These can be used to control external radiation using metasurfaces \cite{holloway2012overview, lin2014dielectric}. Here we consider a singular metasurface. Singularities in plasmonic systems, such as sharp edges or touching points, concentrate the electromagnetic fields even to sub-nanometric volumes yielding huge energy densities \cite{kelly2003optical, prodan2003hybridization, hao2004electromagnetic, nordlander2004plasmon, lu2005nanophotonic, romero2006plasmons, bukasov2007highly, romero2008plasmon, wu2009optical, aubry2010plasmonic, luo2010surface, benz2016single, galiffi2018broadband}. 
 
Transformation optics takes advantage of the coordinate invariance of Maxwell's equations to give a prescription of how the electromagnetic parameters $\epsilon$ and $\mu$ change under geometrical transformations. For the case of two-dimensional conformal transformations, $\epsilon$ and $\mu$ are left unchanged in the plane, which can be exploited for solving complex plasmonics problems by transforming them to a frame where geometry is simpler \cite{Pendry2012}. This is particularly useful when considering systems with singularities \cite{luo2010surface} as they give rise to divergences that cannot be treated exactly with numerical methods. 
 
In this paper we use transformation optics to derive an analytical theory of the optical response of singular plasmonic metasurfaces. In previous works we have applied this framework to the study of metasurfaces with a smooth shape \cite{Kraft2015, Huidobro2017a}. Here we use the conformal transformation introduced in Ref. \cite{pendry2017compacted} for the design of singular metasurfaces, and present an analytical derivation to completely characterize its optical properties.

The text is structured as follows. First, in Section \ref{sec:singularmetasurface} we present the conformal transformation that generates a metasurface with grooves or wedges forming a sharp angle. The external fields incident on the metasurface are considered in Sec. \ref{sec:source}. Next, in Sec. \ref{sec:Fields} we derive the analytical expressions for the electromagnetic fields and absorption cross section in the metasurface. We then introduce in Sec. \ref{sec:conductivitymodel} a flat surface model from which we obtain an effective surface conductivity that allows us to unambiguously determine the optical response of the metasurface through its reflectivity. Finally, the results for the singular metasurface at normal incidence are presented in Sec. \ref{sec:reflection}, and for two cases which break the symmetry in Sec. \ref{sec:breakingsymmetry}: a symmetric metasurface under oblique incidence and an asymmetric metasurface.

\section{\label{sec:singularmetasurface}Creating a singular metasurface}
 
\begin{figure}
\includegraphics[width=1\columnwidth]{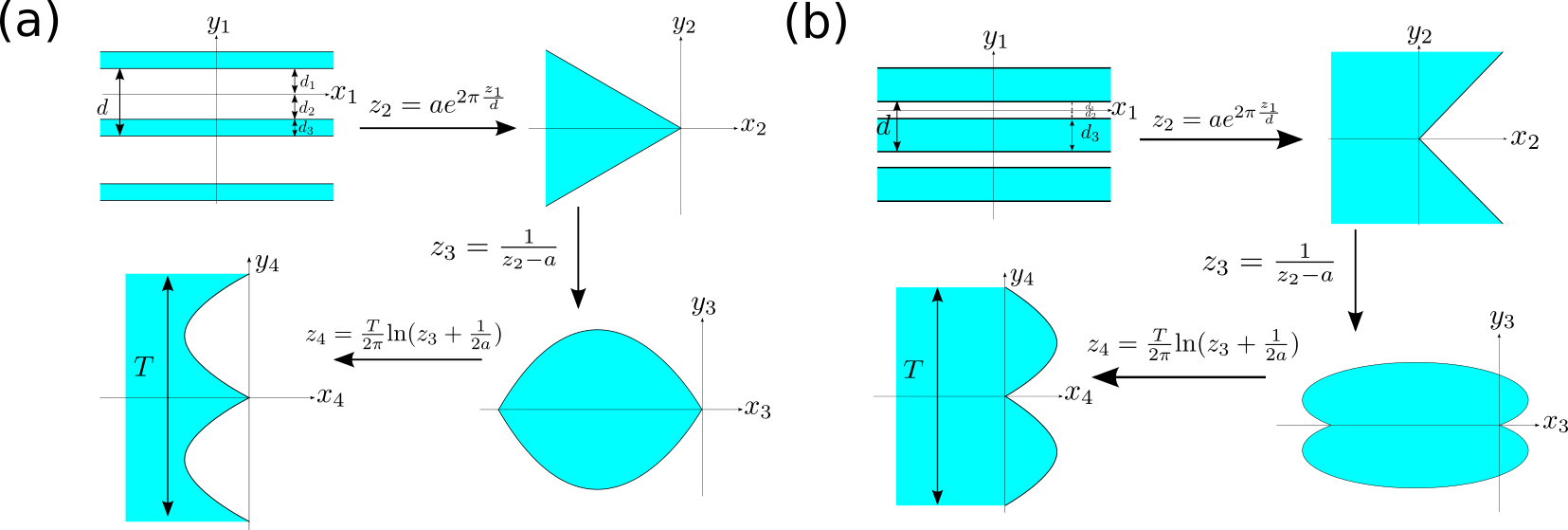}
\centering
\caption{A series of transformations to generate a singular metasurface with (a) concave shape ($d_1+d_2>d_3$) and (b) convex shape ($d_1+d_2<d_3$). The different coordinate frames are labelled as $z_i$, with $z_i=x_i+i y_i$. In the slab frame, $z_1$, the period of the slab array is $d$, $d_1 + d_2$ is the thickness of the air region and $d_3$ is the thickness of each metal slab. This is successively transformed to a single wedge/groove (frame $z_2$), two-touching-circular segments (frame $z_3$) and finally a singular metasurface (frame $z_4$). Note that x-axis and y-axis do not have the same scale. }
\label{Transformation}
\end{figure}

We start by describing the transformations that result in the singular metasurfaces shown in Fig. \ref{Transformation}. Let us first consider an array of metal slabs with periodicity along the vertical direction and translational invariance along the horizontal direction, placed in the slab frame ($z_1=x_1+iy_1$). The period of the array is $d$, the thickness of the slabs is $d_3$ and we take $d_1+d_2$ to be the thickness of the dielectric region. An exponential transformation maps the slab array into either a wedge when $d_1+d_2 > d_3$ [panel (a)], or a groove when $d_1+d_2 < d_3$ [see panel (b)], in frame $z_2$. Then an inverse transformation is carried out to get the two-touching-circular segments shown in the $z_3$ frame. As a last step, a logarithmic transformation is used to generate a surface with a periodic set of sharp wedges/grooves (see $z_4$ frame): these are the singular metasurfaces under consideration here. 
 
On the basis of this series of transformations, a one-step transformation from the slab frame to the metasurface frame can be written as
\begin{equation}
\begin{split}
z_4=\frac{T}{2\pi} \mathrm{ln} \bigg( \frac{1}{a(e^{2\pi z_1 /d}-1)}+\frac{1}{2a} \bigg)
\end{split}
\label{transformation_eq}
\end{equation}
Here, $T$ defines the size of the metasurface by fixing its period, and $a$ is chosen as 0.5 so that the singular point in the metasurface frame is located on the y-axis. As mentioned previously, $d=d_1+d_2+d_3$ is the periodicity in the slab frame, and the choice of $\{d_1, d_2, d_3\}$ determines the shape of the metasurface. For $d_1+d_2=d_3$ the transformation generates a flat surface, while $d_1+d_2 > (<) d_3$ results in concave (convex) singular metasurfaces formed by periodic sharp wedges (grooves). In addition, setting $d_1=d_2$ generates a metasurface that is symmetric with respect to the horizontal axis. 
 
In the following we detail our analytical derivations to calculate the optical response of these singular metasurfaces. Throughout the paper we take the metal to be gold with permittivity approximated by the Drude model, $\varepsilon = 1 - \frac{\omega_p^2}{\omega(\omega+i\Gamma)}$, with plasma frequency $\omega_p = 8.95 $ eV/$\hbar$ and damping $\Gamma = 65.8 $ meV/$\hbar$ \cite{novotny2012principles}.
 
\section{\label{sec:source}Transforming the source}
 
The problem we set out to solve is that of a p-polarized plane wave (magnetic field out of the plane, $H_z$) incident on the singular metasurface, as any more complex wave front can be expressed as a superposition of plane waves. Since the transformation not only transforms the geometry but also the form of the source, we have to derive the representation of the source in the slab frame. For this purpose, we generate an incident wave using a periodic array of magnetic current line sources (``monopoles'') in the right hand-side of the metasurface frame, as depicted in Fig. 2(b). Taking the sources to be at infinity such that their near fields can be safely neglected \cite{pendry1975chain, tretyakov2003analytical}, their radiated field is a  plane wave incident on the surface. We assume that the period of the metasurface is much less than the free space wavelength. When the incident wave impinges on the singular surface, there will be reflected and transmitted waves. These three sorts of waves all participate in the excitation of surface plasmons at the singular surface. In addition to the source currents, there needs to be a sink at infinity to receive the reflected waves, and another sink at minus infinity to receive the transmitted waves. Then, the source in the slab frame can be obtained by recognising that: (i) a magnetic current line is conserved under the transformation and, (ii) sources at $+\infty$ in the metasurface frame are mapped to the point $z_1 = i n d$ while sources at $-\infty$ in the metasurface frame are mapped to $z_1 = i (n+\frac{1}{2})d$ (here $n$ is an integer). Hence, the monopole sources generating the incident wave and the monopoles receiving the reflected waves are located in the air region in the slab frame, while the monopoles receiving the transmitted wave are placed in the metal region [see Fig. 2(a)].
 
We start by writing the magnetic field of the incident, reflected and transmitted waves in the metasurface frame as,
\begin{equation}
\left\{
{\begin{array}{l}
H_z^{inc} = H_0 e^{-i k_{0x} x_4 + i k_{0y} y_4}\\
H_z^{ref} = r H_0 e^{i k_{0x} x_4 + i k_{0y} y_4}\\
H_z^{tra} = t H_0 e^{-i k_{0x}^{'}  x_4 + i k_{0y} y_4}
\end{array}}
\right. \label{eq:source}
\end{equation}
where $H_0$ is the wave amplitude, $r$ and $t$ are reflection and transmission coefficients, $k_{0x}= \sqrt{k_0^2-k_{0y}^2}$ and $k_{0x}^{'}= \sqrt{\varepsilon k_0^2-k_{0y}^2}$.
 
\begin{figure}
\includegraphics[width=0.6\columnwidth]{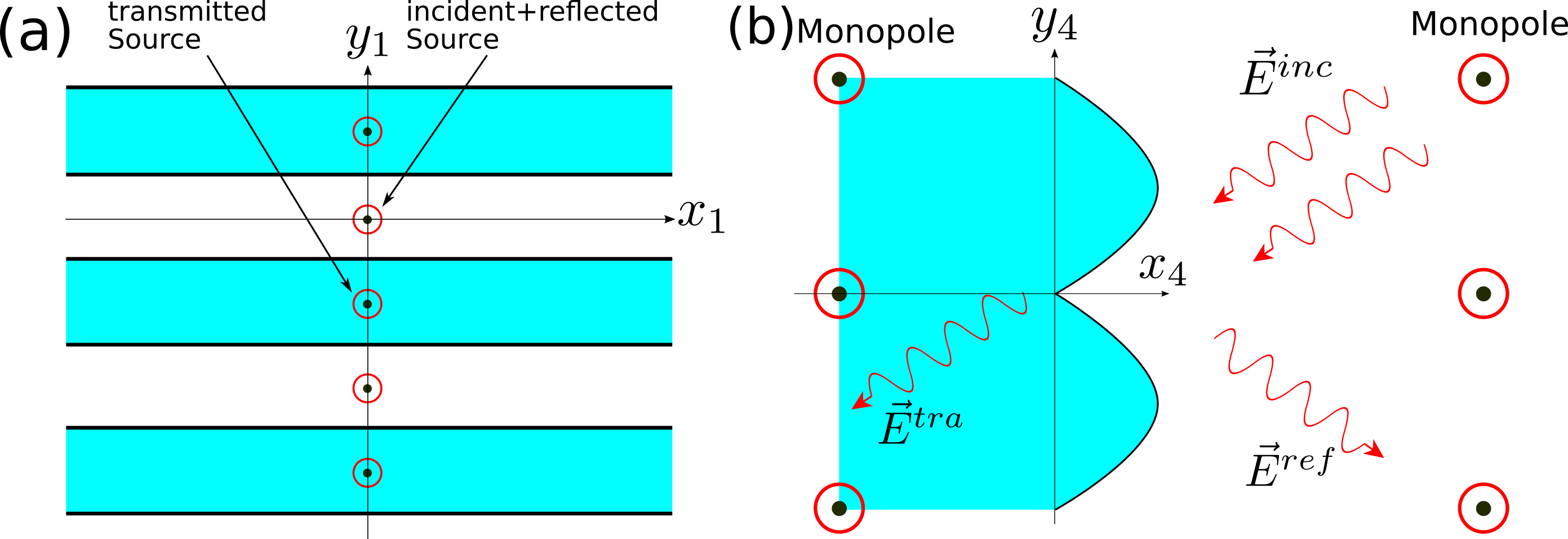}
\centering
\caption{Sketch showing the source field in both frames. In the metasurface frame (b) the source is a plane wave incident on the surface, which we take to be generated by an array of magnetic line currents located at infinity. The source is mapped into an array of magnetic line currents in the slab frame (a). }
\label{Source transformation}
\end{figure}
 
In order to write the source field in the slab frame, we transform Eq. \ref{eq:source} using the mapping Eq. (\ref{transformation_eq}) to obtain,
\begin{equation}
\left\{
{\begin{array}{l}
H_z^{inc} = H_0 \bigg(1+i \frac{k_{0x} T}{2\pi} \mathrm{ln} \big( \frac{\pi}{d} \big) \bigg) +  \int\limits_{-\infty}^{\infty} a_a \frac{e^{-|k_x||y_1|}}{|k_x|} e^{i k_x x_1} d k_x +  \int\limits_{-\infty}^{\infty} a_s \frac{e^{-|k_x||y_1|}}{sgn(y_1)k_x} e^{i k_x x_1} d k_x\\
H_z^{ref} = r H_0 \bigg(1-i \frac{k_{0x} T}{2\pi} \mathrm{ln} \big( \frac{\pi}{d} \big) \bigg) -  \int\limits_{-\infty}^{\infty} r a_a \frac{e^{-|k_x||y_1|}}{|k_x|} e^{i k_x x_1} d k_x +  \int\limits_{-\infty}^{\infty} r a_s \frac{e^{-|k_x||y_1|}}{sgn(y_1)k_x} e^{i k_x x_1} d k_x\\
H_z^{tra} = t H_0 \bigg(1 - i \frac{k_{0x}^{'} T}{2\pi} \mathrm{ln} \big( \frac{\pi}{d} \big) \bigg) - \int\limits_{-\infty}^{\infty} t \frac{k_{0x}^{'}}{k_{0x}}  a_a \frac{e^{-|k_x||y_1|}}{|k_x|} e^{i k_x x_1} d k_x  -  \int\limits_{-\infty}^{\infty} t a_s  \frac{e^{-|k_x||y_1+\frac{d}{2}|}}{sgn(y_1+\frac{d}{2})k_x} e^{i k_x x_1} d k_x
\end{array}}
\right.
\end{equation}
The detailed derivation for these source representations is included in Appendix A. In writing the above equations we have assumed that the period of the metasurface is subwavelength ($T\ll |x_4| \ll \lambda$). Also, we have written the fields as a Fourier series and we have identified a symmetric and an antisymmetric component to the source with amplitudes $a_a = -i \frac{k_{0x}T}{4\pi}H_0$ (anti-symmetric source) and $a_s = \frac{k_{0y}T}{4\pi}H_0$ (symmetric source). Note that we define the symmetry of the modes by considering the $E_x$ component in the slab frame: the anti-symmetric (symmetric) mode has odd (even) symmetry of $E_x(y)$. From the above, we can write the source field in k-space for the antisymmetric mode as,
\begin{equation}
    H_z^a(k_x)=       
        \left\{
        {\begin{array}{lr}
         (1-r)a_a \frac{e^{-|k_x||y|}}{|k_x|},&{ - {d_2} < y < {d_1}} \\
         -t \frac{k_{0x}^{'}}{k_{0x}}  a_a \frac{e^{-|k_x||y+\frac{d}{2}|}}{|k_x|},&{ - {d_2+d_3} < y < {-d_2}}
        \end{array}}
        \right.
\end{equation}
and for the symmetric mode as,
\begin{equation}
       H_z^s(k_x)= \left\{
        {\begin{array}{lr}
         (1+r)a_s \text{sgn}(k_x) \frac{e^{-|k_x||y|}}{\text{sgn}(y)|k_x|},&{ - {d_2} < y < {d_1}} \\
         -t a_s \text{sgn}(k_x) \frac{e^{-|k_x||y+\frac{d}{2}|}}{\text{sgn}(y+\frac{d}{2})|k_x|},&{ - {d_2+d_3} < y < {-d_2}}
        \end{array}}
        \right.
\end{equation}
in which the constant field components are ignored since they do not contribute to the excitation of surface plasmon polaritons. Finally, note that the source representation includes terms $\sim\frac{e^{-|k_x||y|}}{|k_x|}$, which is just the Fourier transformation of a Hankel function in the quasi-static limit \cite{chew1995waves}. Indeed, a Hankel function is the source representation of a line current, which further confirms our source representation.  
 
\section{\label{sec:Fields}Electromagnetic fields and absorption cross section}
 
\subsection{\label{sec:dispersion}Surface plasmons dispersion relation}
Once we have the source fields, we can calculate the excited field components. When a SPP mode ($H_z \sim e^{i k_x x}$) on the boundary between metal and air is excited, the total field distribution in the slab frame can be written as   
\begin{equation}
H_z(k_x)=\left\{{\begin{array}{lr}
(1-r)a_a \frac{e^{-|k_x||y|}}{|k_x|} + (1+r)a_s \frac{e^{-|k_x||y|}}{\text{sgn}(y)k_x}+ {b_ +}e^{-|k_x|y} + {b_ -}e^{|k_x|y},&{ -d_2 < y < d_1}\\
-t \frac{k_{0x}^{'}}{k_{0x}}  a_a \frac{e^{-|k_x||y+\frac{d}{2}|}}{|k_x|} -t a_s \frac{e^{-|k_x||y+\frac{d}{2}|}}{\text{sgn}(y+\frac{d}{2})k_x} + {c_ +}e^{-|k_x|y} + {c_ -}e^{|k_x|y},&{ - ({d_2} + {d_3}) < y <  - {d_2}}
\end{array}} \right.
\end{equation}
Here, $b_+$, $b_-$, $c_+$ and $c_-$ are the excited mode amplitudes, which have to be determined from the boundary conditions. This requires matching the tangent components of the fields, $H_z$ and $E_x$, at $y=d_1$, $y=-d_2$ and $y=-(d_2+d_3)$ \cite{luo2010surface} and details are given in Appendix B. 
 
\begin{figure}[h]
\includegraphics[width=0.8\columnwidth]{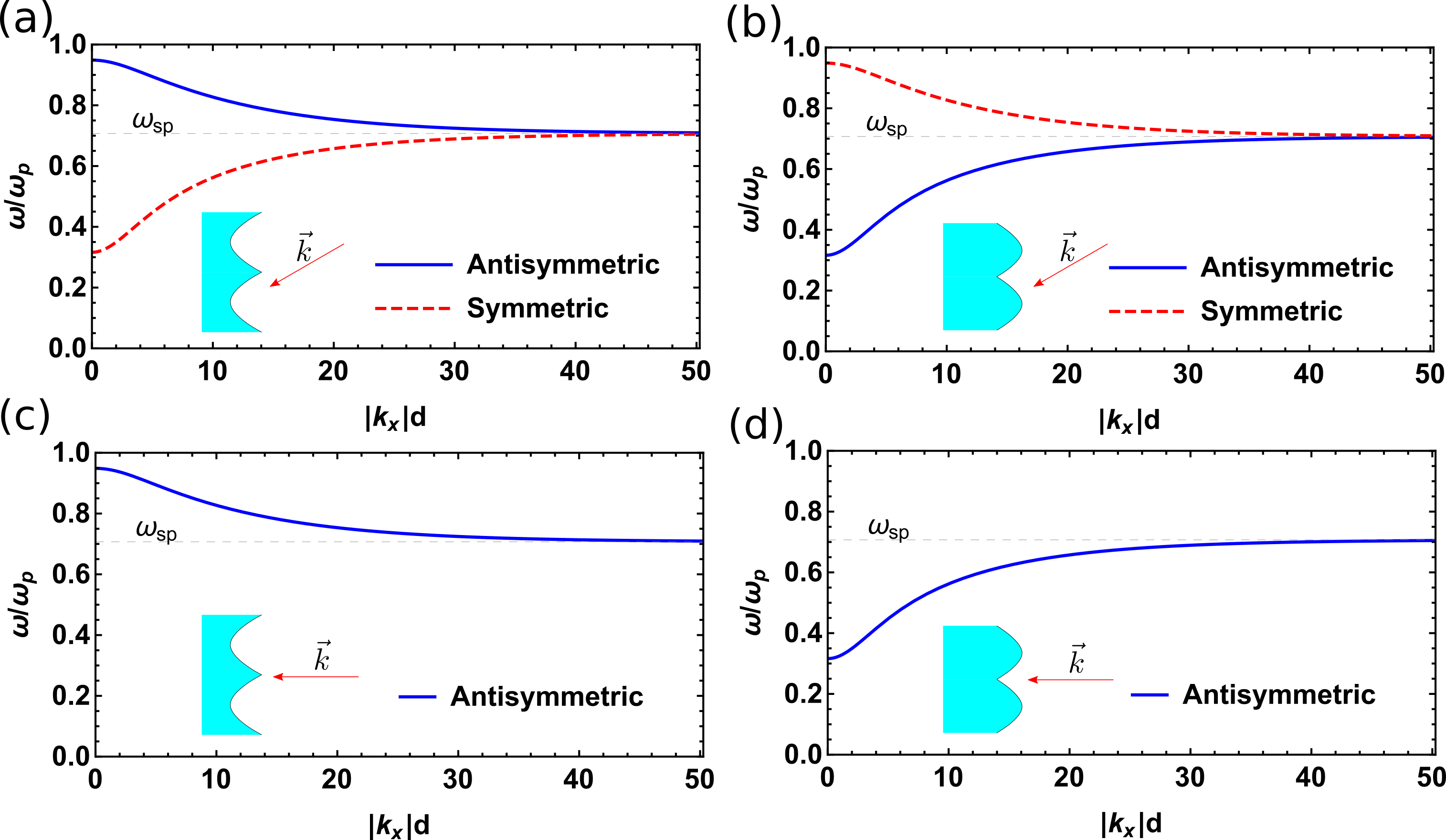}
\centering
\caption{Dispersion relations of surface plasmons in singular metasurfaces. (a), (c) SPP dispersion relation in a wedge singular metasurface at oblique (a) and normal incidence (c). The parameters are $d_3=0.1d$, $d_1=d_2=(d-d_3)/2$.(b), (d) SPP dispersion relation in a groove singular metasurface at oblique (b) and normal incidence (d). The parameters are $d_3=0.9d$, $d_1=d_2=(d-d_3)/2$.}
\label{Dispersion relation}
\end{figure}
 
Once the excited field mode amplitudes have been determined (see Eqs. B2-B5), the dispersion relation of surface plasmons can be obtained by looking at the poles of these coefficients. Neglecting the pole $k_x = 0$ as it is a branch point which corresponds to a localized virtual excitation rather than SPPs \cite{aubry2010broadband, aubry2010conformal}, we concentrate on the plasmon pole. From the amplitude of the anti-symmetric mode (Eqs. B2 and B3), we obtain
\begin{equation}
\begin{split}
(\varepsilon-1)(e^{|k_{x}|(d_1+d_2)} - e^{|k_{x}|d_3}) + (\varepsilon+1)(e^{|k_{x}|d}-1)=0
\end{split}
\end{equation}
While for the symmetric mode (Eqs. B4 and B5) we have,
\begin{equation}
\begin{split}
(\varepsilon-1)(e^{|k_{x}|(d_1+d_2)} - e^{|k_{x}|d_3}) - (\varepsilon+1)(e^{|k_{x}|d}-1)=0
\end{split}
\end{equation}

Figure \ref{Dispersion relation} shows the calculated dispersion relations for the singular wedge (a, c) and groove (b, d) metasurfaces. For the general case, at an oblique incidence, both the symmetric and antisymmetric bands can be excited, as shown in panels (a) and (b). On the other hand, at normal incidence only the antisymmetric band is excited [panels (c) and (d)]. It should be noted that the bands have lower and upper cut-off frequencies different from 0 and $\omega_p$.
For the wedge (groove) metasurface, the symmetric (antisymmetric) band spans a frequency range $\omega_{c1}\le\omega<\omega_{sp}$ while the antisymmetric (symmetric band) spans $\omega_{sp}<\omega\le \omega_{c2}$. Here, $\omega_{sp}$ is the surface plasmon frequency, and $\omega_{c(1,2)}$ are the lower and upper cut-off frequencies, which were shown to be $\omega_{c1}= \omega_{p} \sqrt{\frac{\theta}{2\pi}}$ and $\omega_{c2} = \omega_{p} \sqrt{\frac{2\pi-\theta}{2\pi}}$, where $\theta = 2\pi \frac{d_3}{d}$ for the wedge case ($d_1+d_2>d_3$), $\theta = 2\pi \frac{d_1+d_2}{d}$ for the groove case ($d_1+d_2<d_3$) \cite{luo2010surface}. The appearance of these cut-offs is a result of plasmon hybridization in the infinite periodic array of the slab frame \cite{luo2010surface}. Note that when analyzing the dispersion relation, we have assumed a lossless metal, while we use a finite damping for the rest of the paper. 
 
In the following we focus on the groove singular metasurface, for which the antisymmetric band excited at normal incidence exists below $\omega_{sp}$, where the Drude model gives a more accurate description of the metal.

\subsection{\label{sec:Fieldsrealspace}Electric and magnetic fields in real space}
 
Next, we calculate the electromagnetic fields in real space. We write the SPP mode field distribution by taking the Fourier transform of the excited terms in the magnetic field given by Eq. (6), 
\begin{equation}
    H_z(x,y) = \int\limits_{-\infty}^\infty H_z(k_x,y)e^{i{k_x}x}d{k_x} = 2\pi i \text{Res}\left[H_z(k_x,y)e^{i{k_x}x} \right]|_{k_x=k_{px}}
\end{equation}    
where the residue theorem is applied at the plasmon pole, $k_x = k_{px}$. Also, from Eq. (6) we have $H_z (k_x,y) = {b_ +}e^{-|k_x|y} + {b_ -}e^{|k_x|y}$ in the dielectric region and  $H_z (k_x,y) = {c_ +}e^{-|k_x|y} + {c_ -}e^{|k_x|y}$ in the metal region. This yields the following field distribution in the slab frame: 
\begin{equation}
    H_z(x,y)= \left\{
        {\begin{array}{lr}
        i 2\pi a ({\Gamma _ + }{e^{ - {\sqrt{k_{px}^2}}y}} + {\Gamma _ - }{e^{{\sqrt{k_{px}^2}}y}})e^{i{k_{px}}|x|},&{ - {d_2} < y < {d_1}}\\
        i 2\pi a ({\Lambda _ + }{e^{ - {\sqrt{k_{px}^2}}y}} + {\Lambda _ - }{e^{{\sqrt{k_{px}^2}}y}})e^{i{k_{px}}|x|},&{ - {d_2+d_3} < y < -{d_2}}
        \end{array}}
        \right. 
\end{equation}
where $a$ stands for $a_{(a,s)}$ for the antisymmetric/symmetric mode and $|k_x|$ has been written as $\sqrt{k_x^2}$ in the complex integration. All the field coefficients ($\Gamma_+$, $\Gamma_-$, $\Lambda_+$ and $\Lambda_-$) are given in Appendix B (Eqs. B6-B9). The electric field can be derived from Maxwell's equations using $E_x= \frac{i}{\omega \varepsilon} \frac{\partial{H_z}}{\partial{y}}$,
\begin{equation}
    E_x(x,y) = \left\{
        {\begin{array}{lr}
        \frac{2\pi a \sqrt{k_{px}^2}}{\omega \varepsilon_0}({\Gamma _ + }{e^{ - {\sqrt{k_{px}^2}}y}} - {\Gamma _ - }{e^{{\sqrt{k_{px}^2}}y}})e^{i{k_{px}}|x|},&{ - {d_2} < y < {d_1}}\\
        \frac{2\pi a \sqrt{k_{px}^2}}{\omega \varepsilon_0 \varepsilon}({\Lambda _ + }{e^{ - {\sqrt{k_{px}^2}}y}} - {\Lambda _ - }{e^{{\sqrt{k_{px}^2}}y}})e^{i{k_{px}}|x|},&{ - {d_2+d_3} < y < -{d_2}}
        \end{array}}
        \right.
\end{equation}
and $E_y = -\frac{i}{\omega \varepsilon} \frac{\partial{H_z}}{\partial{x}}$,
\begin{equation}
    E_y(x,y) = \left\{
        {\begin{array}{lr}
        i \text{sgn}(x) \frac{2\pi a k_{px}}{\omega \varepsilon_0}({\Gamma _ + }{e^{ - {\sqrt{k_{px}^2}}y}} + {\Gamma _ - }{e^{{\sqrt{k_{px}^2}}y}})e^{i{k_{px}}|x|},&{ - {d_2} < y < {d_1}}\\
        i \text{sgn}(x) \frac{2\pi a k_{px}}{\omega \varepsilon_0 \varepsilon}({\Lambda _ + }{e^{ - {\sqrt{k_{px}^2}}y}} + {\Lambda _ - }{e^{{\sqrt{k_{px}^2}}y}})e^{i{k_{px}}|x|},&{ - {d_2+d_3} < y < -{d_2}}
        \end{array}}
        \right.
\end{equation}
 
\begin{figure}
\includegraphics[width=0.6\columnwidth]{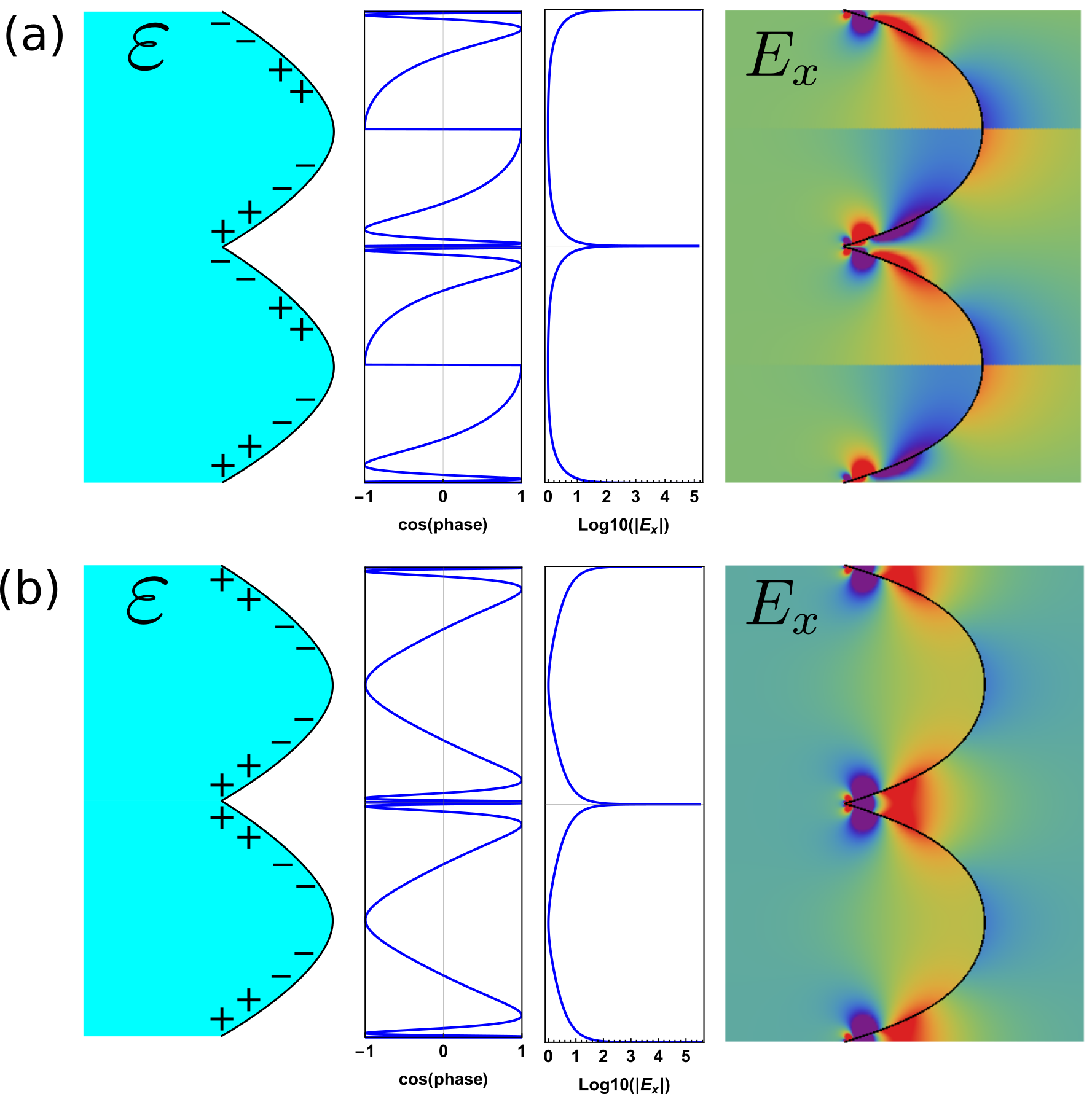}
\centering
\caption{Mode plot of $E_x$ in the metasurface frame for (a) anti-symmetric mode at $\omega=0.6\omega_p$ and (b) symmetric mode at $\omega=0.8\omega_p$. From left column to right column are: charge distribution (sketch), phase and amplitude along the singular surface, and field distribution in the unit cell. The parameters are: $d_3=0.9d$, $d_1=d_2=(d-d_3)/2$. In the color map, red stands for positive values of the field magnitude and blue for negative ones.}
\label{modeplot}
\end{figure}
 
Once we have the fields in the slab frame, the fields in the metasurface frame can be calculated by mapping them following the rules of transformation optics\cite{Ward1996,Pendry2006}. The obtained mode profile is plotted in Fig. \ref{modeplot} for the antisymmetric (a) and symmetric (b) modes at two frequencies of choice below and above the surface plasmon frequency. We show a sketch of the charge distribution for each mode (antibonding and bonding, respectively), together with the calculated phase and amplitude of the $E_x$ component along the singular surface, and a field plot for $E_x(x,y)$ in the metasurface frame. It is clear the phase oscillates very rapidly, and the in-plane electric field diverges at the singularity. This can be understood from the compression of an infinite dimension hidden at the singular point \cite{pendry2017compacted}, making the derivative of $H_z$ with respect to $x$ and $y$ to be infinite at the singularity. On the other hand, due to its invariance, the magnetic field keeps a finite value in all the frames shown in Fig. \ref{Transformation}, and in particular in the singular metasurface frame.  
 
The presence of loss will attenuate the field in any real system. In this case, the amplitude of the excited SPP wave in the slab frame will be attenuated as it travels along the slab and away from the sources. As was shown previously, there is a critical angle of the groove/wedge where the enhancement by compression and attenuation by loss are balanced \cite{luo2010surface}. The electric field converges for $\theta \leq \theta_c$, while it diverges otherwise, where $\theta_c = \mathrm{Im}[ \mathrm{ln} \left( \frac{\varepsilon-1}{\varepsilon+1} \right) ]$ for the lower band, and $\theta_c = -\mathrm{Im}[ \mathrm{ln} \left( \frac{1-\varepsilon}{\varepsilon+1} \right) ]$ for the upper band. The geometrical parameters chosen in Fig. \ref{modeplot} ($d_3=0.9d$, $d_1=d_2=(d-d_3)/2$) yield an angle at the singularity of $\theta = 2\pi \frac{Min(d_1+d_2,d_3)}{d} = 0.2 \pi$. At the frequencies of choice, $\omega=0.6\omega_p$ for the antisymmetric mode and $\omega=0.8\omega_p$ for the symmetric one, the critical angles are $\theta_c = 0.03$ and $\theta_c = 0.04$, respectively. Therefore, the electric field at singular point of both anti-symmetric and symmetric modes diverges despite the presence material losses, as shown in the right column of Fig. \ref{modeplot}.
 
\subsection{\label{sec:dissipation}Energy dissipation by SPPs}
Once the field distribution is obtained, we calculate the energy dissipated by the excited SPP mode in the slab frame. Dissipation is due to loss in the metal, so we calculate the absorbed power as the following integral on the slab volume, 
\begin{equation}
\begin{split}
    P^{(a,s)}_{abs} &= \int\limits_{slab} \frac{1}{2} \omega \varepsilon_0 \mathrm{Im}[\varepsilon] |E|^2 dxdy\\
    &= \frac{4\pi^2|a|^2|k_{px}|^2 \mathrm{Im}[\varepsilon]}{\omega \varepsilon_0 |\varepsilon|^2} (\frac{|\Lambda_{(a,s)+}|^2}{-2\mathrm{Re}[\sqrt{k_{px}^2}]} (e^{2 \mathrm{Re}[\sqrt{k_{px}^2}]d_2}-e^{2 \mathrm{Re}[\sqrt{k_{px}^2}](d_2+d_3}) \\
    &+ \frac{|\Lambda_{(a,s)-}|^2}{2\mathrm{Re}[\sqrt{k_{px}^2}]}(e^{-2 \mathrm{Re}[\sqrt{k_{px}^2}]d_2}-e^{-2 \mathrm{Re}[\sqrt{k_{px}^2}](d_2+d_3)}) \big) \frac{1}{\mathrm{Im}[k_{px}]}
\end{split}
\end{equation}
where the coefficients $\Lambda_{(a,s)\pm}$ stand for the antisymmetric and symmetric mode coefficients given in Eqs. (B7) and (B9). Using $\Lambda_{a\pm}$ ($\Lambda_{s\pm}$)  yields the power absorbed by the antisymmetric (symmetric) mode. Finally, the absorption cross section of the structure can be obtained by normalizing to the input energy on the system in one period,  
\begin{equation}
    \sigma^{(a,s)}_{abs} = \frac{P^{(a,b)}_{abs}}{\frac{1}{2} \sqrt{\frac{\mu_0}{\varepsilon_0}} H_0^2 T \cos{\theta_{in}}},
\end{equation}
where $\theta_{in}$ is incident angle of the plane wave and $H_0$ is the wave amplitude. Thus, we have for the anti-symmetric mode, 
\begin{equation}
\begin{split}
    \sigma_{abs}^a
    &= \frac{k_0 T \cos{\theta_{in}}}{2}\frac{|k_{px}|^2 \mathrm{Im}[\varepsilon]}{|\varepsilon|^2} (\frac{|\Lambda_{a+}|^2}{-2\mathrm{Re}[\sqrt{k_{px}^2}]} (e^{2 \mathrm{Re}[\sqrt{k_{px}^2}]d_2}-e^{2 \mathrm{Re}[\sqrt{k_{px}^2}](d_2+d_3}) \\
    &+ \frac{|\Lambda_{a+}|^2}{2\mathrm{Re}[\sqrt{k_{px}^2}]} ( e^{-2 \mathrm{Re}[\sqrt{k_{px}^2}]d_2}-e^{-2 \mathrm{Re}[\sqrt{k_{px}^2}](d_2+d_3)}) \big) \frac{1}{\mathrm{Im}[k_{px}]}
\end{split}
\end{equation}
and for the symmetric one,
\begin{equation}
\begin{split}
    \sigma_{abs}^s
    &= \frac{k_0 T \sin^2{\theta_{in}}}{2\cos{\theta_{in}}}\frac{|k_{px}|^2 \mathrm{Im}[\varepsilon]}{|\varepsilon|^2} (\frac{|\Lambda_{s+}|^2}{-2\mathrm{Re}[\sqrt{k_{px}^2}]} (e^{2 \mathrm{Re}[\sqrt{k_{px}^2}]d_2}-e^{2 \mathrm{Re}[\sqrt{k_{px}^2}](d_2+d_3}) \\
    &+ \frac{|\Lambda_{s+}|^2}{2\mathrm{Re}[\sqrt{k_{px}^2}]} ( e^{-2 \mathrm{Re}[\sqrt{k_{px}^2}]d_2}-e^{-2 \mathrm{Re}[\sqrt{k_{px}^2}](d_2+d_3)}) \big) \frac{1}{\mathrm{Im}[k_{px}]}
\end{split}
\end{equation}

\subsection{\label{sec:blunt}Metasurface with blunt singularities: From a continuous to a discrete spectrum}

\begin{figure}
\includegraphics[width=0.6\columnwidth]{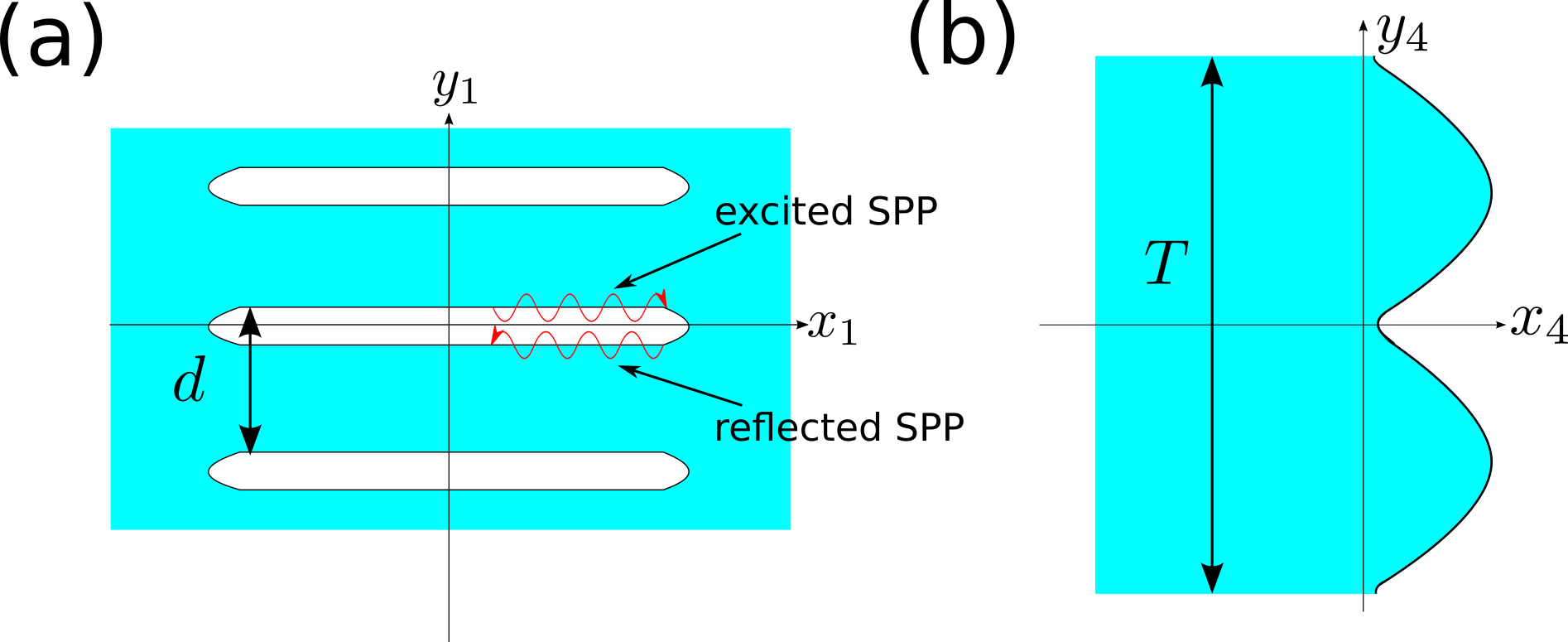}
\centering
\caption{Metasurface with blunt singularities. A periodic array of truncated slabs in the slab frame (a) maps to a metasurface with blunt singularities in the metasurface frame (b). }
\label{Transformation_discretespectrum}
\end{figure}

Any fabricated metasurface will not show a perfect singularity but a blunt one, so we now move on to treat blunt singularities \cite{luo2012broadband}. A metasurface with rounded singularities maps into a truncated slab array (or truncated cavity array for the groove geometry), as shown in Fig. \ref{Transformation_discretespectrum}. This has an important consequence: when the slabs/cavities are not infinite, the excited SPP modes travelling along the slab will be reflected at its terminals, as depicted in Fig. \ref{Transformation_discretespectrum}(a), resulting in a quantization of the SPP modes supported by the slab structure. Hence, the continuous spectrum of a singular metasurface turns into a discrete spectrum when the singularities are blunt. 
 
In order to calculate the field distribution and energy dissipation in the truncated slab/cavity arrays we first need to calculate the reflection coefficient of SPPs at the slab/cavity terminal. For conciseness, in the following we will refer only to the cavity array shown in Fig. \ref{Transformation_discretespectrum} (a) and we point out here that the same derivation can be applied to the truncated slab array. The radius of the rounded singularity in the metasurface frame determines the length of the cavities in the slab frame, $L$, as well as the exact shape of the terminal. In order to calculate the reflection coefficient of the SPP modes at the end of the cavity we assume that the terminal is a flat vertical air/metal interface. Then we consider the field of the SPP modes ($H_z^{sp}$) in the cavity, $-L/2<x_1<L/2$, and the field ($H_z^{out}$) in the region outside the cavity, $x_1<-L/2$ and $x_1>L/2$, and impose the continuity condition of the tangential fields and the power flow at $x_1=L/2$ \cite{chandran2012metal, gordon2006light}. This yields the equations,
\begin{eqnarray}
(1+r_{sp})H_z^{sp} &=& H_z^{out} \label{eq:contH}\\
(1-r_{sp})E_y^{sp} &=& E_y^{out} \label{eq:contE}\\
\int\limits_{-(d_2+d_3)}^{d_1}(1-r_{sp}^*)(1+r_{sp})E_y^{sp*} H_z^{sp} dy &=& \int\limits_{-(d_2+d_3)}^{d_1} E_y^{out*} H_z^{out} dy \label{eq:contS}
\end{eqnarray}
where, $H_z^{sp}$ and $E_y^{sp}$ are given in Sec. \ref{sec:Fieldsrealspace}.
 
In the region outside the periodic cavity array we expand the magnetic field as a series of Bloch waves,
\begin{equation}
\begin{split}
H_z^{out} = \sum_{g} h(g) e^{i g y}
\end{split}
\end{equation}
where $g=n \frac{2\pi}{d}$ and $n$ is an integer. The coefficients $h(g)$ can be expressed as 
\begin{equation}
\begin{split}
h(g) &= \frac{1}{d} \int\limits_{-(d_2+d_3)}^{d_1} H_z^{out} e^{-i g y} dy 
= \frac{1+r_{sp}}{d} \int\limits_{-(d_2+d_3)}^{d_1} H_z^{sp} e^{-i g y} dy 
= \frac{1+r_{sp}}{d} I(g) \label{eq:hg}
\end{split}
\end{equation}
where we have made use of Eq. \ref{eq:contH} and we have named the integral in the last equality as $I(g)$. On the other hand, we can derive the electric field in this region as $E_y^{out} = - \frac{i}{\omega \varepsilon \varepsilon_0} \frac{\partial H_z^{out}}{\partial x}$. This yields
\begin{equation}
\begin{split}
E_y^{out} =  \frac{1}{\omega \varepsilon \varepsilon_0} \sum_{g} \sqrt{\varepsilon k_0^2 - g^2} h(g) e^{i g y}
\end{split}
\end{equation}
where we have used $h(g) \propto e^{i \sqrt{\varepsilon k_0^2 - g^2}x}$ and $\varepsilon$ should be replaced by $1$ for the wedge system. Replacing the expression for $h(g)$, Eq. \ref{eq:hg}, in the above equation we obtain
\begin{equation}
\begin{split}
E_y^{out} = \frac{1+r_{sp}}{\omega \varepsilon \varepsilon_0 d} \sum_{g} \sqrt{\varepsilon k_0^2 - g^2} I(g) e^{i g y}
\end{split}
\end{equation}
Substituting the expressions for $H_z^{out}$ and $E_y^{out}$ into Eq. \ref{eq:contS} and after some algebra we arrive to
\begin{equation}
\begin{split}
\frac{1-r_{sp}}{1+r_{sp}} = \frac{1}{ \omega \varepsilon \varepsilon_0 d} \frac{\sum_{g}  \left( \sqrt{\varepsilon k_0^2 - g^2} \right) |I(g)|^2}{\int\limits_{-(d_2+d_3)}^{d_1} E_y^{sp} H_z^{sp^*} dy} \equiv G
\end{split} \label{eq:G}
\end{equation}
where we take $k_0 = 0$ in the quasi-static limit. From Eq. \ref{eq:G}, we obtain the reflection coefficient as
\begin{equation}
    r_{sp} = \frac{1-G}{1+G} \label{eq:rsp}
\end{equation}

Since in the quasistatic limit radiative loss is small, the amplitude of the reflection coefficient is $|r_{sp}|\approx 1$ (which is indeed confirmed from our calculations using Eq. \ref{eq:rsp}). Hence, we consider only the reflection phase, $r_{sp} = e^{i \phi}$, and write the field of the SPP mode in the cavity of length $L$,
\begin{equation}
    H_z(x,y)= \left\{
        {\begin{array}{lr}
        \begin{split}
            & i 2\pi a ({\Gamma_+}e^{-|k_{px}|y}+ {\Gamma_-}e^{|k_{px}|y}) \\
            & \times (e^{i k_{px} |x|} + e^{-i k_{px} |x|+ i k_{px} L + i\phi}) \frac{1}{1 \mp e^{i k_{px} L + i \phi}}
        \end{split}
        ,&{ - {d_2} < y < {d_1}}\\
        \begin{split}
            & i 2\pi a ({\Lambda_+}e^{-|k_{px}|y}+ {\Lambda_-}e^{|k_{px}|y}) \\
            & \times (e^{i k_{px} |x|} + e^{-i k_{px} |x|+ i k_{px} L + i\phi}) \frac{1}{1 \mp e^{i k_{px} L + i \phi}}
        \end{split}
        ,&{ - {d_2+d_3} < y < -{d_2}}
        \end{array}}
        \right.
\end{equation}
in which $\mp$ stands for $-$ for the anti-symmetric source, and $+$ for the symmetric source. From $H_z$ the electric field is obtained using $E_x(x,y) = \frac{i}{\omega \varepsilon} \frac{\partial{H_z}}{\partial{y}} $ and $E_y(x,y) = -\frac{i}{\omega \varepsilon} \frac{\partial{H_z}}{\partial{x}} $.
 
Finally, the absorption cross section of the blunt singular metasurface is derived from the electric field following the same procedure detailed in Sec. \ref{sec:dissipation}. We have for the antisymmetric mode,
\begin{equation}
\begin{split}
    \sigma_{abs}^a
    &= \frac{k_0 T \cos{\theta_{in}} }{2} \frac{|k_{px}|^2 \mathrm{Im}[\varepsilon]}{|\varepsilon|^2} \frac{1}{|1-e^{i k_{px} L + i \phi}|^2} 
    \bigg[ (\frac{|\Lambda_{a+}|^2}{-2 \mathrm{Re}[\sqrt{k_{px}^2}]} (e^{2 \mathrm{Re}[\sqrt{k_{px}^2}]d_2}-e^{2 \mathrm{Re}[\sqrt{k_{px}^2}](d_2+d_3)}) \\
    & + \frac{|\Lambda_{a-}|^2}{2 \mathrm{Re}[\sqrt{k_{px}^2}]} (e^{-2 \mathrm{Re}[\sqrt{k_{px}^2}]d_2}-e^{-2 \mathrm{Re}[\sqrt{k_{px}^2}](d_2+d_3)})) \frac{1-e^{-2\mathrm{Im}[k_{px}]L}}{\mathrm{Im}[k_{px}]} \\
    &- (\frac{\Lambda_{a+}^* \Lambda_{a-}}{i 2 \mathrm{Im}[\sqrt{k_{px}^2}]} (e^{-i 2 \mathrm{Im}[\sqrt{k_{px}^2}]d_2} - e^{-i 2 \mathrm{Im}[\sqrt{k_{px}^2}](d_2+d_3)}) \\
    &+ \frac{\Lambda_{a+} \Lambda_{a-}^*}{-i 2 \mathrm{Im}[\sqrt{k_{px}^2}]} (e^{i 2 \mathrm{Im}[\sqrt{k_{px}^2}]d_2} - e^{i 2 \mathrm{Im}[\sqrt{k_{px}^2}](d_2+d_3)})) \frac{2 e^{-\mathrm{Im}[k_{px}]L}}{\mathrm{Re}[k_{px}]}(\sin(\mathrm{Re}[k_{px}]L+\phi)-\sin(\phi)) \bigg]
\end{split}
\end{equation}
and for the symmetric one, 
\begin{equation}
\begin{split}
    \sigma_{abs}^s
    &= \frac{k_0 T \sin^2{\theta_{in}} }{2 \cos{\theta_{in}}} \frac{|k_{px}|^2 \mathrm{Im}[\varepsilon]}{|\varepsilon|^2} \frac{1}{|1+e^{i k_{px} L + i \phi}|^2} 
    \bigg[ (\frac{|\Lambda_{s+}|^2}{-2 \mathrm{Re}[\sqrt{k_{px}^2}]} (e^{2 \mathrm{Re}[\sqrt{k_{px}^2}]d_2}-e^{2 \mathrm{Re}[\sqrt{k_{px}^2}](d_2+d_3)}) \\
    & + \frac{|\Lambda_{s-}|^2}{2 \mathrm{Re}[\sqrt{k_{px}^2}]} (e^{-2 \mathrm{Re}[\sqrt{k_{px}^2}]d_2}-e^{-2 \mathrm{Re}[\sqrt{k_{px}^2}](d_2+d_3)})) \frac{1-e^{-2\mathrm{Im}[k_{px}]L}}{\mathrm{Im}[k_{px}]} \\
    &- (\frac{\Lambda_{s+}^* \Lambda_{s-}}{i 2 \mathrm{Im}[\sqrt{k_{px}^2}]} (e^{-i 2 \mathrm{Im}[\sqrt{k_{px}^2}]d_2} - e^{-i 2 \mathrm{Im}[\sqrt{k_{px}^2}](d_2+d_3)}) \\
    &+ \frac{\Lambda_{s+} \Lambda_{s-}^*}{-i 2 \mathrm{Im}[\sqrt{k_{px}^2}]} (e^{i 2 \mathrm{Im}[\sqrt{k_{px}^2}]d_2} - e^{i 2 \mathrm{Im}[\sqrt{k_{px}^2}](d_2+d_3)})) \frac{2 e^{-\mathrm{Im}[k_{px}]L}}{\mathrm{Re}[k_{px}]}(\sin(\mathrm{Re}[k_{px}]L+\phi)-\sin(\phi)) \bigg]
\end{split}
\end{equation}
 
\section{\label{sec:conductivitymodel}Modeling a singular metasurface with an effective surface conductivity}
 
\begin{figure}
\includegraphics[width=0.6\columnwidth]{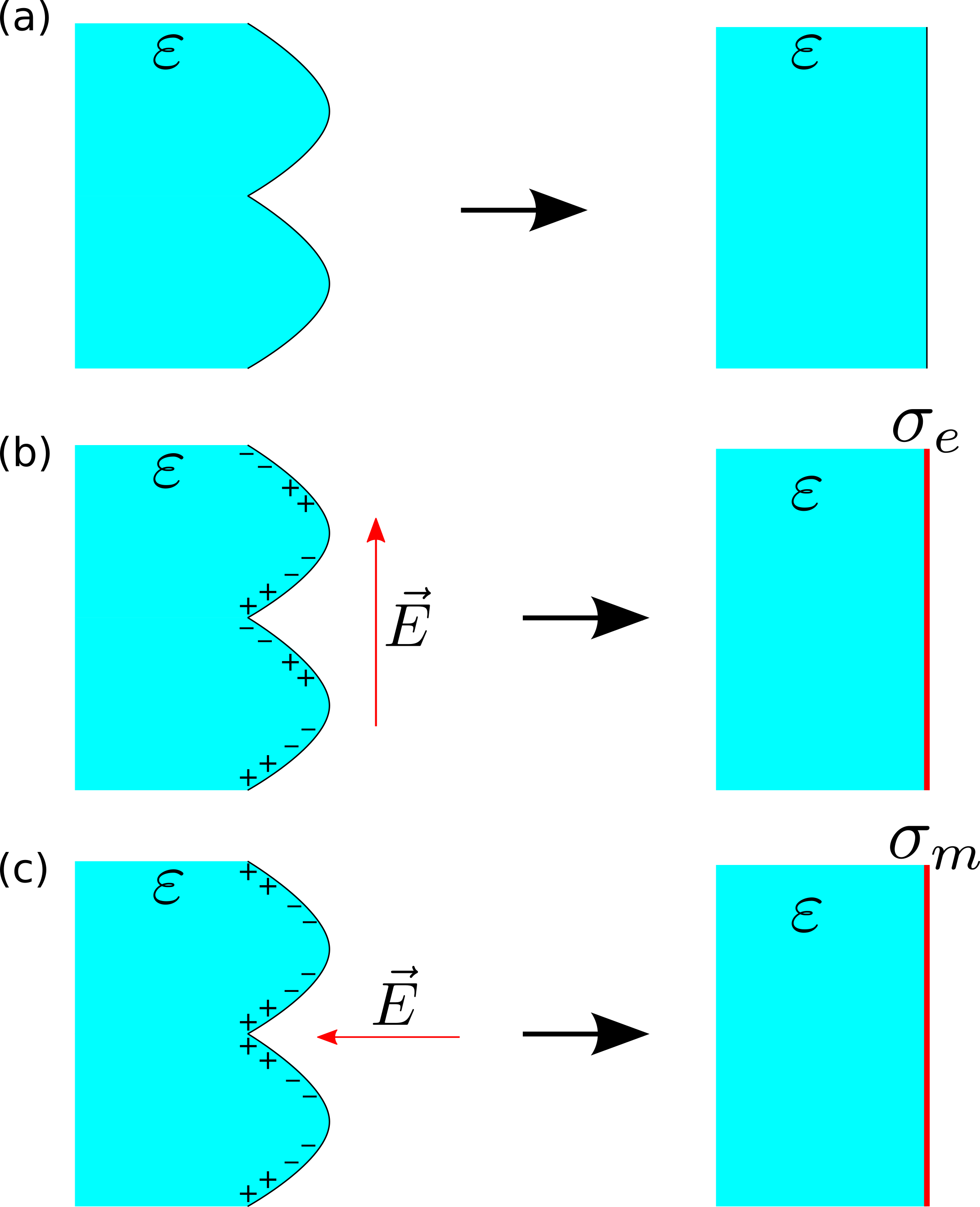}
\centering
\caption{Flat surface model: (a) when SPP is not excited, the singular metasurface behaves as a flat surface; (b) when the anti-symmetric SPP mode is excited, the singular metasurface is modeled as a flat surface with an electric surface conductivity $\sigma_e$; (c) when the symmetric SPP mode is excited, the singular metasurface is modeled as a flat surface with a magnetic surface conductivity $\sigma_m$.}
\label{Flat surface model}
\end{figure}
 
The next step in our analytical treatment is to calculate reflection off the singular metasurface. For this purpose, we have developed an effective surface conductivity model. The model is based on the assumption of subwavelength periodicity and is illustrated in Fig. \ref{Flat surface model}. If there is no SPP mode excited on the metasurface, an incident wave will not see it and as a consequence the singular surface behaves effectively as a flat surface (a). If on the other hand SPPs are excited, energy will be dissipated at the metasurface, which we model through an effective surface conductivity. Due to their different charge distributions at the metasurface, the antisymmetric SPP mode will induce an electric surface current (b), while the symmetric SPP mode will induce a magnetic surface current (c). Hence, we model the antisymmetric mode with an effective electric conductivity, $\sigma_e=\sigma_{er} + i \sigma_{ei}$, and the symmetric mode with an effective magnetic conductivity, $\sigma_m=\sigma_{mr} + i \sigma_{mi}$

Let us first consider the antisymmetric mode. The effective electric surface current generated by the excited SPP modes yields the discontinuity of the tangential magnetic field, while the tangential electric field is continuous,
\begin{eqnarray}
    E_y^{inc}+E_y^{ref}-E_y^{t} &=& 0     \\
    H_z^{inc}+H_z^{ref}-H_z^{t} &=& -\sigma_e E_y^{loc}
\end{eqnarray}
Here, $E_y^{loc}$ is the local tangent electric field in the homogenized system, $E_y^{loc} = \frac{1}{2}(E_y^{inc}+E_y^{ref}+E_y^{tra})$. From the first of the above equations we have $t = \frac{k_{0x} \varepsilon}{k_{0x}^{'}}(1-r)$, so we write 
\begin{equation}
    \begin{split}
      E_y^{loc} &= -(1-r)\frac{k_{0x}}{\omega \varepsilon_0} H_0  \label{eq:Eyloc}
    \end{split}
\end{equation}
where we see that the amplitude of the anti-symmetric mode is $\propto 1-r$. In fact, using the proportionality between $t$ and $1-r$ for this mode, we find that the mode coefficients for this mode can be re-written as $\Lambda_{a\pm}=(1-r)\Lambda'_{a\pm}$ and $\Gamma_{a\pm}=(1-r)\Gamma'_{a\pm}$, where the primed coefficients are independent of $r$ and $t$. Detailed expressions for these normalized coefficients are given in Appendix C.

Similarly, for the symmetric mode we have an effective magnetic surface current which yields,
\begin{eqnarray}
    E_y^{inc}+E_y^{ref}-E_y^{t} &=& -\sigma_m H_z^{loc}     \\
    H_z^{inc}+H_z^{ref}-H_z^{t} &=& 0
\end{eqnarray}
where the local tangent magnetic field in the homogenized system is $H_z^{loc} = \frac{1}{2}(H_z^{inc}+H_z^{ref}+H_z^{tra})$. In this case we have $t=1+r$, so
\begin{equation}
    \begin{split}
      H_z^{loc} &= (1+r) H_0  \label{eq:Hzloc}
    \end{split}
\end{equation}
and the amplitude of the symmetric mode is $\propto 1+r$. Normalized mode coefficients can be defined in this case by writing $\Lambda_{s\pm}=(1+r)\Lambda'_{s\pm}$ and $\Gamma_{s\pm}=(1+r)\Gamma'_{s\pm}$. We give detailed expressions in Appendix C. 
 
From the above derivation it is clear that the anti-symmetric mode amplitude is proportional to the local electric field $E_y^{loc}$, while the symmetric mode amplitude is proportional to the local magnetic field $H_z^{loc}$. This justifies the introduction of two kinds of surface conductivities, electric and magnetic ones, in our model in order to mimic energy dissipation by the excited SPP wave. With this, the complex singular metasurface has been simplified as an easy boundary value problem. The reflection coefficient of the singular metasurface can be written straightforwardly as
{\scriptsize
\begin{equation}
\begin{split}
        r=\frac{-4 \varepsilon  \sigma_m + 4 \sigma_e Z_0^2 \cos \theta_{in} \sqrt{\varepsilon -\sin ^2\theta_{in}}-\sigma_e \sigma_m Z_0 \sqrt{\varepsilon -\sin ^2\theta_{in}}+\varepsilon  \sigma_e \sigma_m Z_0 \cos \theta_{in}-4 Z_0 \sqrt{\varepsilon -\sin ^2\theta_{in}}+4 \varepsilon  Z_0 \cos \theta_{in}}{4 \varepsilon  \sigma_m + 4 \sigma_e Z_0^2 \cos \theta_{in} \sqrt{\varepsilon -\sin ^2\theta_{in}}+\sigma_e \sigma_m Z_0 \sqrt{\varepsilon -\sin ^2\theta_{in}}+\varepsilon  \sigma_e \sigma_m Z_0 \cos\theta_{in}+4 Z_0 \sqrt{\varepsilon -\sin ^2\theta_{in}}+4 \varepsilon  Z_0 \cos\theta_{in}}
\end{split}
\label{reflection}
\end{equation}
}
where $\theta_{in}$ is the angle of incidence and $Z_0 = \sqrt{\frac{\mu_0}{\varepsilon_0}}$ is the impedance of free space. Note that for $\sigma_{(e,m)}=0$,  Eq. \ref{reflection} reduces to the reflection coefficient for a flat surface.
 
The problem then reduces to finding the effective conductivities. We first derive their real parts by using energy conservation, as this is the term that takes away energy. The energy absorbed in the metasurface, $\sigma_{abs} P_{inc}$, must equal energy dissipated by the excited SPP, so we write for the symmetric and antisymmetric modes,
\begin{equation}
\left\{
\begin{split}
\frac{1}{2} \sigma_{er} |E_y^{loc}|^2 T &= \sigma_{abs}^{a} P_{inc}\\
\frac{1}{2} \sigma_{mr} |H_z^{loc}|^2 T &= \sigma_{abs}^{s} P_{inc}
\end{split}
\right.
\label{energy conservation}
\end{equation}
where $P_{inc}=\frac{1}{2} Z_0 H_0^2 T \cos{\theta_{in}}$. By substituting the expression of the local tangent fields, Eqs. \ref{eq:Eyloc} and \ref{eq:Hzloc}, we arrive to 
\begin{equation}
\left\{
\begin{split}
    \sigma_{er}  &= \frac{\omega^2 \varepsilon_0^2}{k_{0x}^2} \frac{\sigma^a_{abs}}{|1-r|^2} Z_0 \cos{\theta_{in}} \\
    \sigma_{mr}  &= \frac{\sigma_{abs}^{s}}{|1+r|^2} Z_0 \cos{\theta_{in}}
\end{split}
\right.
\end{equation}
Now, noting that for the antisymmetric mode $t = \frac{k_{0x} \varepsilon}{k_{0x}^{'}}(1-r)$, we have that $\sigma^a_{abs}\propto|1-r|^2$ (see Appendix C for detailed derivations), and we can eliminate the reflection coefficient in the equation for $\sigma_{er}$. Similarly, for the symmetric mode we have $t=1+r$, so $\sigma^s_{abs}\propto|1+r|^2$ and we can also eliminate the reflection coefficient in the equation for $\sigma_{mr}$. We can then write,
\begin{equation}
\left\{ 
\begin{split}
    \sigma_{er}  &= \frac{\sigma_{abs}^{a'}}{Z_0} = \sigma_{abs}^{a'} \sigma_{e0} \\
    \sigma_{mr}  &= \sigma_{abs}^{s'}Z_0 \sin^2{\theta_{in}} = \sigma_{abs}^{s'} \sigma_{m0} \sin^2{\theta_{in}} \label{eq:sigmar}
    \end{split}
\right.
\end{equation}
where $\sigma_{e0} = Z_0^{-1}$ and $\sigma_{m0} = Z_0$ are the free space electric and magnetic conductivity and we have introduced the intrinsic absorption cross sections (denoted with primes), which do not depend on the incident angle. Hence, the real parts of the effective electric and magnetic conductivities are given by the intrinsic absorption cross sections of the anti-symmetric and symmetric modes, respectively. These read as, 
\begin{equation}
\begin{split}
\sigma_{abs}^{(a,s)'} &= \frac{k_0 T}{2} \frac{|k_{px}|^2 \mathrm{Im}[\varepsilon]}{|\varepsilon|^2}
    (\frac{|\Lambda_{(a,s)+}^{'}|^2}{-2\mathrm{Re}[\sqrt{k_{px}^2}]} (e^{2 \mathrm{Re}[\sqrt{k_{px}^2}]d_2}-e^{2 \mathrm{Re}[\sqrt{k_{px}^2}](d_2+d_3}) \\
    &+ \frac{|\Lambda_{(a,s)-}^{'}|^2}{2\mathrm{Re}[\sqrt{k_{px}^2}]} ( e^{-2 \mathrm{Re}[\sqrt{k_{px}^2}]d_2}-e^{-2 \mathrm{Re}[\sqrt{k_{px}^2}](d_2+d_3)}) \big) \frac{1}{\mathrm{Im}[k_{px}]}
\end{split} \label{eq:sigap}
\end{equation}
where all the normalized excited field coefficients are given in Appendix C.
 
Eqs. \ref{eq:sigmar}-\ref{eq:sigap} determine the real parts of the electric and magnetic conductivities unambiguously. That is, $\sigma_{er}$ and $\sigma_{mr}$ are just functions of the frequency and independent of $r$ and $t$. Furthermore, it should be noted that the electric and magnetic conductivities are defined in the frequency ranges where the antisymmetric and symmetric modes are supported, respectively. For a symmetric metasurface, the symmetric and antisymmetric modes exist over different frequency ranges as we discussed in Sec. \ref{Dispersion relation}. Specifically, we may write for the electric conductivity,
\begin{equation}
    \sigma_{er} = \left\{
        {\begin{array}{lr}
         \sigma_{abs}^{a'} \sigma_{e0}, &{\omega_{c1} < \omega < \omega_{sp}}\\
        0, &{other}
        \end{array}}
        \right.
\label{electric surface conductivity}
\end{equation}
where $\omega_{c1} < \omega < \omega_{sp}$ corresponds to the frequency range when the anti-symmetric mode is excited. Likewise, for the magnetic conductivity we have
\begin{equation}
    \sigma_{mr} = \left\{
        {\begin{array}{lr}
         \sigma_{abs}^{s'} \sigma_{m0} \sin^2{\theta_{in}}, &{\omega_{sp} < \omega < \omega_{c2}}\\
        0, &{other}
        \end{array}}
        \right.
\label{magnetic surface conductivity}
\end{equation}
where $\omega_{sp} < \omega < \omega_{c2}$ corresponds to the frequency range when the symmetric mode is excited.
 
Finally, in order to fully determine the conductivities, we also need their imaginary parts. These can be obtained through Kramers-Kronig relations as the conductivity must satisfy causality \cite{jackson1999classical, dressel2005electrodynamics}. The imaginary part is thus calculated using
\begin{equation}
\begin{split}
\sigma_{(e,m)i} = -\frac{1}{\pi}P\int\limits_{-\infty}^{\infty}\frac{\sigma_{(e,m)r}(s)}{s-\omega} ds = \frac{1}{\pi}  P \int\limits_{-\infty}^{\infty}\mathrm{ln}\big|s-\omega\big|\frac{d \sigma_{(e,m)r}(s)}{ds} ds
\end{split}
\label{Kramers-Kronig}
\end{equation}
Using the above equation the complex surface conductivities are fully determined and the reflection coefficient is finally obtained by substituting $\sigma_{e}$ and $\sigma_{m}$ into Eq. \ref{reflection}. Furthermore, we note that Eqs. \ref{eq:sigmar} and \ref{Kramers-Kronig} also hold for the metasurface with blunt singularities with the appropriate expressions for $\sigma_{abs}^{(a,b)'}$. These are given in Appendix \ref{sec:appAbsblunt}.
 
\section{\label{sec:reflection} Reflection spectrum of the singular metasurface}
 
\begin{figure}
\includegraphics[width=0.5\columnwidth]{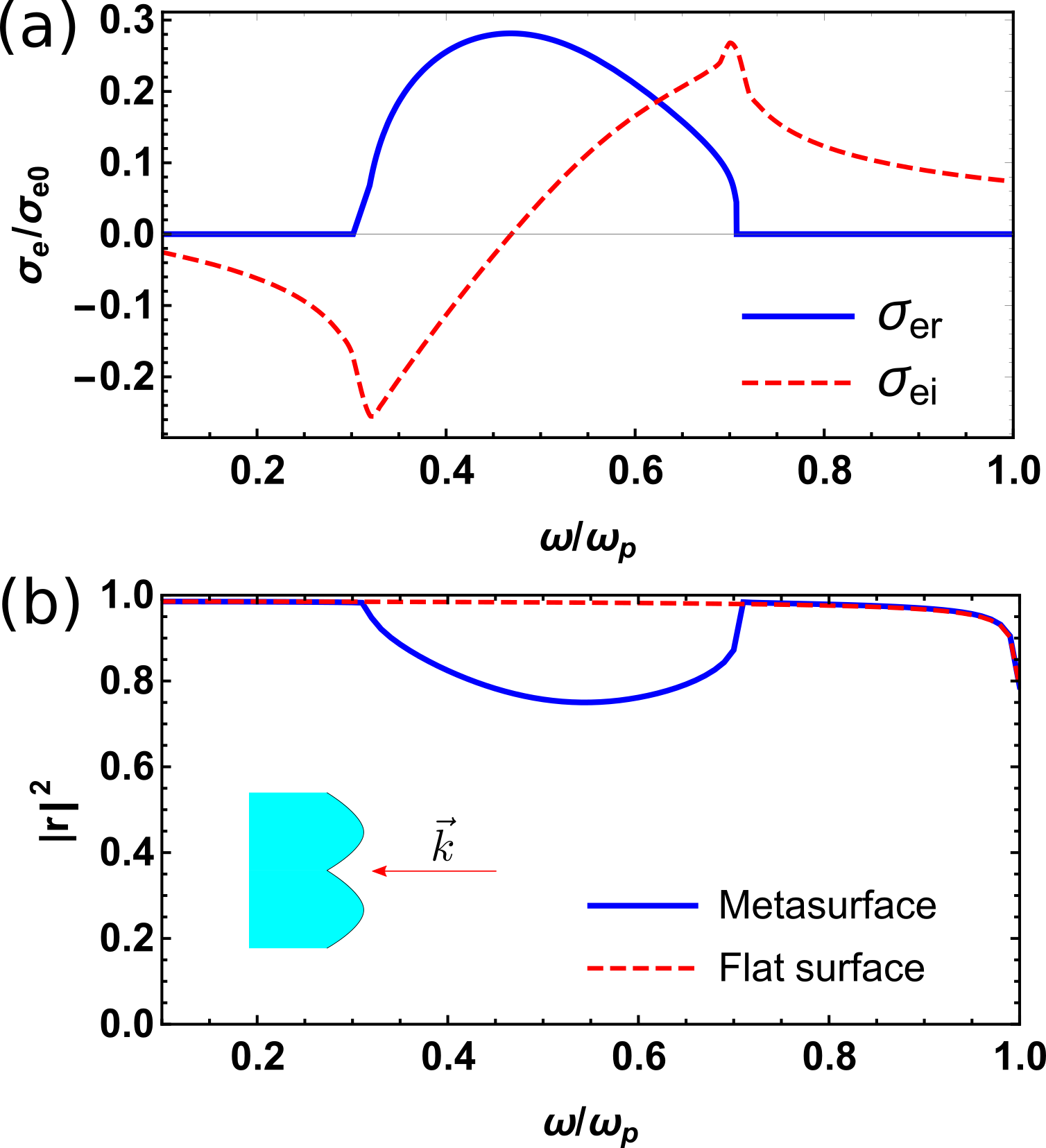}
\centering
\caption{Continuous spectrum of a singular metasurface at normal incidence. (a) Effective electric surface conductivity: real (solid blue line) and imaginary (dashed red) parts. (b) Reflectivity of the metasurface (blue solid line) and of a flat surface with the same permittivity (red dashed). The parameters taken are $T=10$ nm, $d_3=0.9d$, $d_1=d_2=(d-d_3)/2$.}
\label{continuous spectrum}
\end{figure}

Using the analytical framework presented in the previous sections, we now discuss the spectrum of a singular groove metasurface of period $T=10$ nm. The metasurface is defined with the parameters $d_3=0.9d$, $d_1=d_2=(d-d_3)/2$, such that it is symmetric with respect to $y=0$. We first consider a normally incident plane wave as source, such that only the anti-symmetric band is excited. Correspondingly, in this scenario the metasurface is modelled by only an effective electric conductivity. The calculated conductivity is presented in Fig. \ref{continuous spectrum} (a). Its real part, shown as a solid blue line, is non-zero only within the anti-symmetric band (between $\omega_{c1}$ and $\omega_{sp}$), as given by Eq. \ref{electric surface conductivity}. Outside the band, no SPPs are excited so the real part of electric surface conductivity is zero, and the metasurface acts effectively as a flat surface without a surface current. The imaginary part of the conductivity is plotted as a dashed red line. In this case, $\sigma_{ei}$ is non-zero also outside the band, where $\sigma_{er}=0$. This is necessary to satisfy Kramers-Kronings relations and represents a phase shift of the reflected wave at the metasurface. 
 
From the effective conductivity we obtain the reflection coefficient and we plot the reflectivity ($|r|^2$) in panel (b). The solid blue line corresponds to the calculated reflectivity of the singular metasurface, and the red dashed line to the reflectivity of a flat metal surface calculated from Fresnel coefficients. It is clear that outside of the SPP band the optical response of the singular metasurface is the same as that of a flat surface. On the other hand, between $\omega_{c1}$ and $\omega_{sp}$ the reflectivity presents a continuous spectrum where the reflectivity is smaller than 1. This corresponds to the excitation the anti-symmetric SPP band. Note that we only present analytical results here as full wave simulations using commercial software cannot calculate the spectrum of an exactly singular metasurface because the electric field diverges at the singularity.

\begin{figure}
\includegraphics[width=0.5\columnwidth]{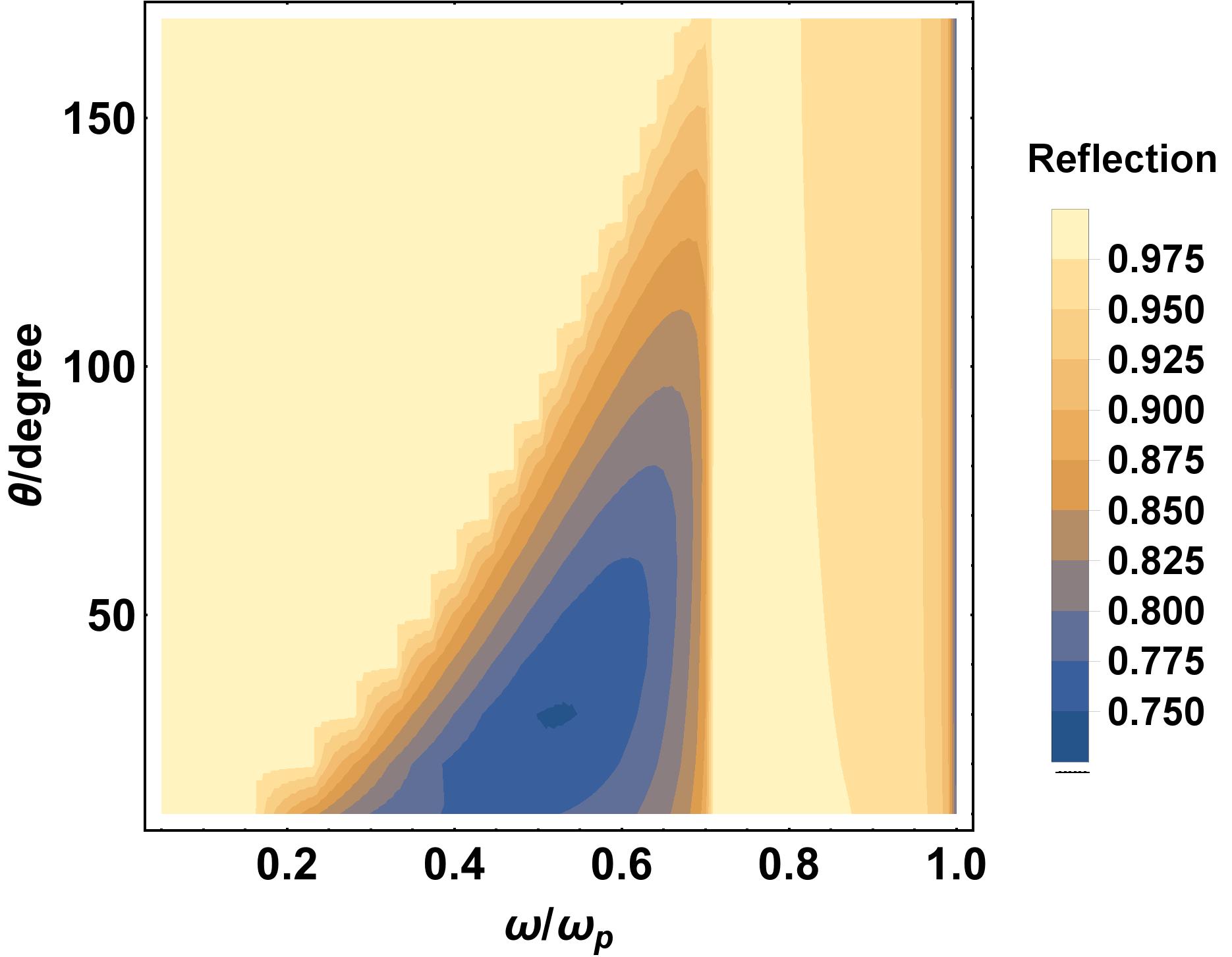}
\centering
\caption{Reflectivity as a function of frequency, $\omega$, and angle of the singular grooves, $\theta$, for a singular metasurface of period $T=10$ nm. }
\label{reflection_parameterplot_thetaVSomega}
\end{figure}
 
\begin{figure}
\includegraphics[width=0.5\columnwidth]{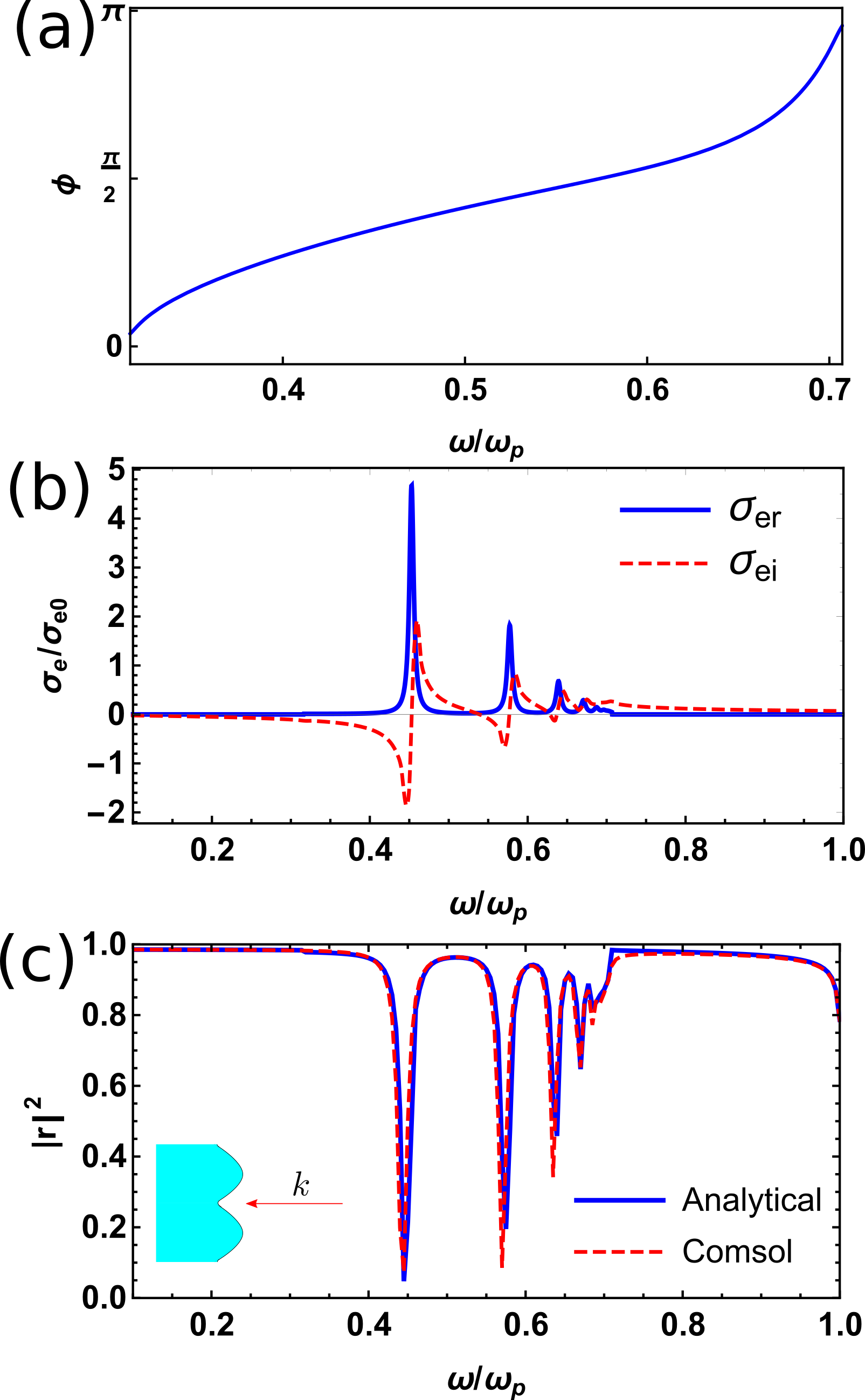}
\centering
\caption{Spectrum of a blunt singular metasurface at normal incidence, showing a discrete rather than a continuum. (a) Phase picked-up by SPPs at the terminal of the truncated cavity. (b) Effective electric surface conductivity, real (solid blue line) and imaginary (red dashed) parts. (c) Reflectivity: analytical (solid blue line) and numerical (red dashed). The metasurface parameters are the same as in Fig. \ref{continuous spectrum}, and the truncation length is $L=d$.}
\label{discrete spectrum}
\end{figure}
 
Next we explore the dependence of the normal incidence reflectivity spectrum on the sharpness of the singularity. Figure \ref{reflection_parameterplot_thetaVSomega} shows the reflectivity as a function of frequency and of the singularity angle, $\theta = 2\pi\frac{d_1+d_2}{d}$. The blue region corresponds to lower values of reflectivity, where the antisymmetric SPP mode is excited, below $\omega_{sp}$. The low reflectivity band is broad for small angles, and it becomes narrower as $\theta$ gets closer to $180^{\circ}$. The reason for this is that the cut-off frequency for this band, $\omega_{c1} = \omega_{p} \sqrt{\frac{\theta}{2\pi}}$, approaches $\omega_{sp} = \frac{\omega_{p}}{\sqrt{2}}$. On the other hand, we can see that the maximum reflection is reached at a finite angle, $\sim30^{\circ}$.

As discussed previously, real singular metasurfaces will present blunt singularities, which results in a discrete rather than a continuum spectrum. Figure \ref{discrete spectrum} presents results for a metasurface with the same parameters as in Fig. \ref{continuous spectrum} but with blunt singularities. Panel (a) shows the phase, $\phi$, acquired by SPPs reflected off the blunt edge, which we calculate in the slab frame for SPPs reflecting at the truncated end of the cavity. From the calculated reflection phase we obtain the electric surface conductivity through Eqs. \ref{electric surface conductivity}, which is plotted in panel (b). For frequencies within the anti-symmetric band, the conductivity develops resonances which result from the quantization of SPP modes in the periodic array of finite cavities. Then, using Eq. \ref{reflection}, we calculate the corresponding reflectivity [see panel (c), solid blue line]. The reflectivity presents a discrete set of peaks, in contrast to the continuous spectrum of the singular metasurface. We also present the reflectivity obtained from full wave simulations (using the commercial finite element method solver Comsol Multiphysics) as a red dashed line, which shows an excellent agreement with the analytical one. This confirms the validity of our analytical modelling.

\section{\label{sec:breakingsymmetry}Breaking the symmetry}
 
In Sec. \ref{sec:reflection} we have considered a symmetric metasurface illuminated with a normally incident plane wave, such that the anti-symmetric band is excited below the surface plasmon frequency while the symmetric band (above $\omega_{sp}$) is dark. In this Section we discuss how both bands can be excited by breaking the symmetry. Two ways of breaking the symmetry will be considered. First, we study a symmetric metasurface under oblique incidence, such that the source breaks the symmetry and both bands can be excited. Second, we consider an asymmetric metasurface. Different from the first case, an asymmetry in the geometry mixes the anti-symmetric and symmetric modes so that they become coupled with each other.  
 
\subsection{\label{sec:obliqueinc}Symmetric metasurface under oblique incidence}
\begin{figure}
\includegraphics[width=0.5\columnwidth]{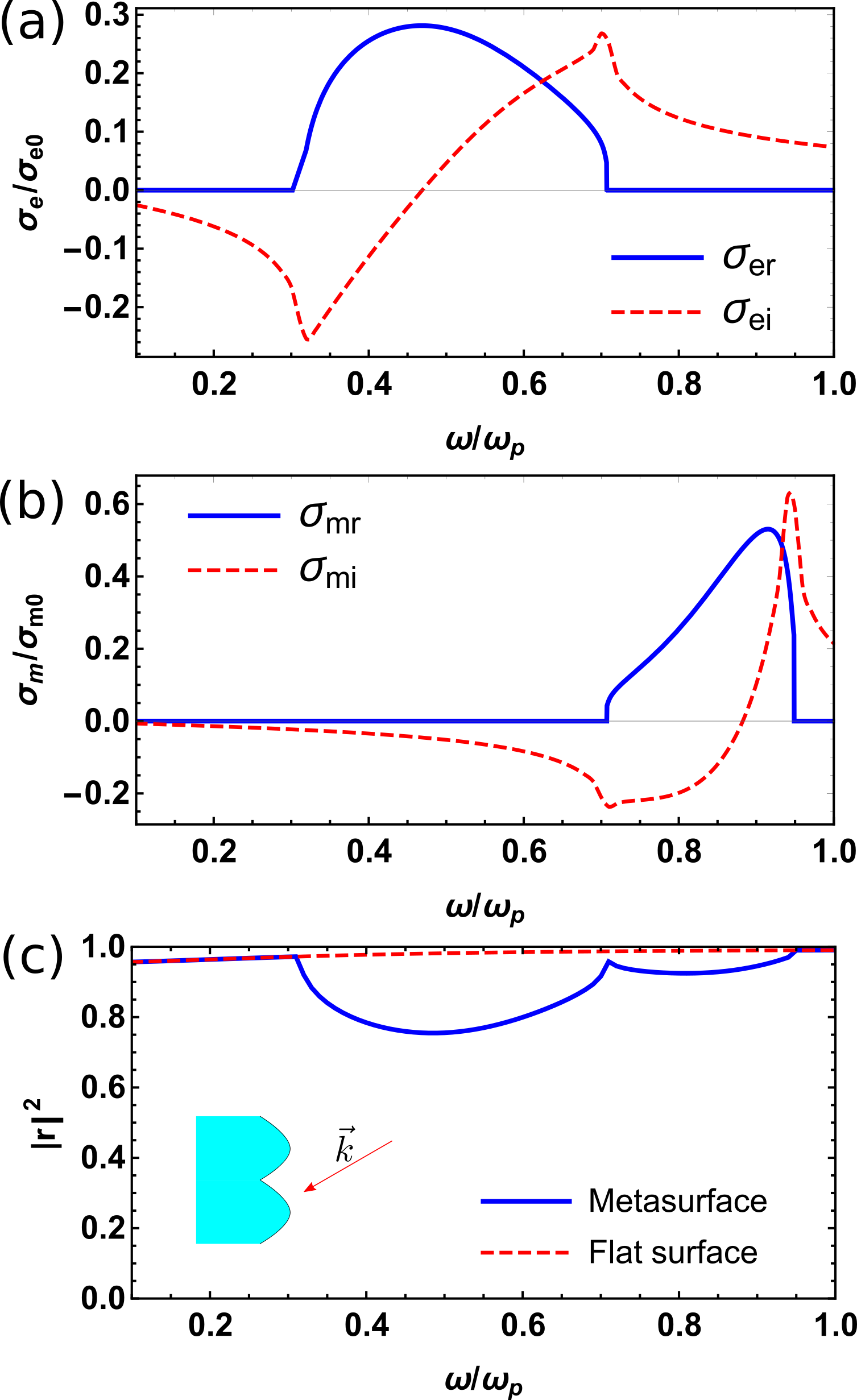}
\centering
\caption{Continuous spectrum of the singular metasurface at oblique incidence ($\theta_{in} = 0.4\pi$). (a,b) Effective electric (magnetic) surface conductivity: real (solid blue line) and imaginary (dashed red) parts. (c) Reflectivity of the metasurface (blue solid line) and of a flat surface with the same permittivity (red dashed). The parameters for the metasurface are the same as in Fig. \ref{continuous spectrum}. }
\label{continuous_spectrum_obliqueincidence}
\end{figure}

When the incident field is coming at an oblique angle of incidence, the source is no longer symmetric with respect to the $y=0$ plane. As a consequence, the dark mode becomes bright and both the anti-symmetric and the symmetric bands are excited. The results obtained with our analytical model for the singular metasurface studied above but under oblique incidence (incident angle $\theta_{in} = 0.4\pi$) are shown in Fig. \ref{continuous_spectrum_obliqueincidence}. Since both anti-symmetric and symmetric modes are excited, there will be both an electric surface conductivity $\sigma_e$ and a magnetic surface conductivity $\sigma_m$. As discussed above, for the groove metasurface the lower band ($\omega_{c1}<\omega<\omega_{sp}$) is the anti-symmetric mode, and its energy dissipation is modelled as a complex electric surface conductivity $\sigma_e$, which is shown in panel (a). Since the conductivity only depends on the intrinsic absorption cross section of SPPs, and not on the incidence angle, it is the same as for the normal incidence case considered above. On the other hand, the upper band is the symmetric mode ($\omega_{sp}<\omega<\omega_{c2}$), which is equivalent to a magnetic surface conductivity $\sigma_m$, shown in panel (b). Finally, panel (c) shows the metasurface reflectivity at oblique incidence. Different from Fig. \ref{continuous spectrum}, where only the lower band is excited, we see how under oblique incidence there are two continuous bands corresponding to the excitation of the anti-symmetric and symmetric modes.

\begin{figure}
\includegraphics[width=1\columnwidth]{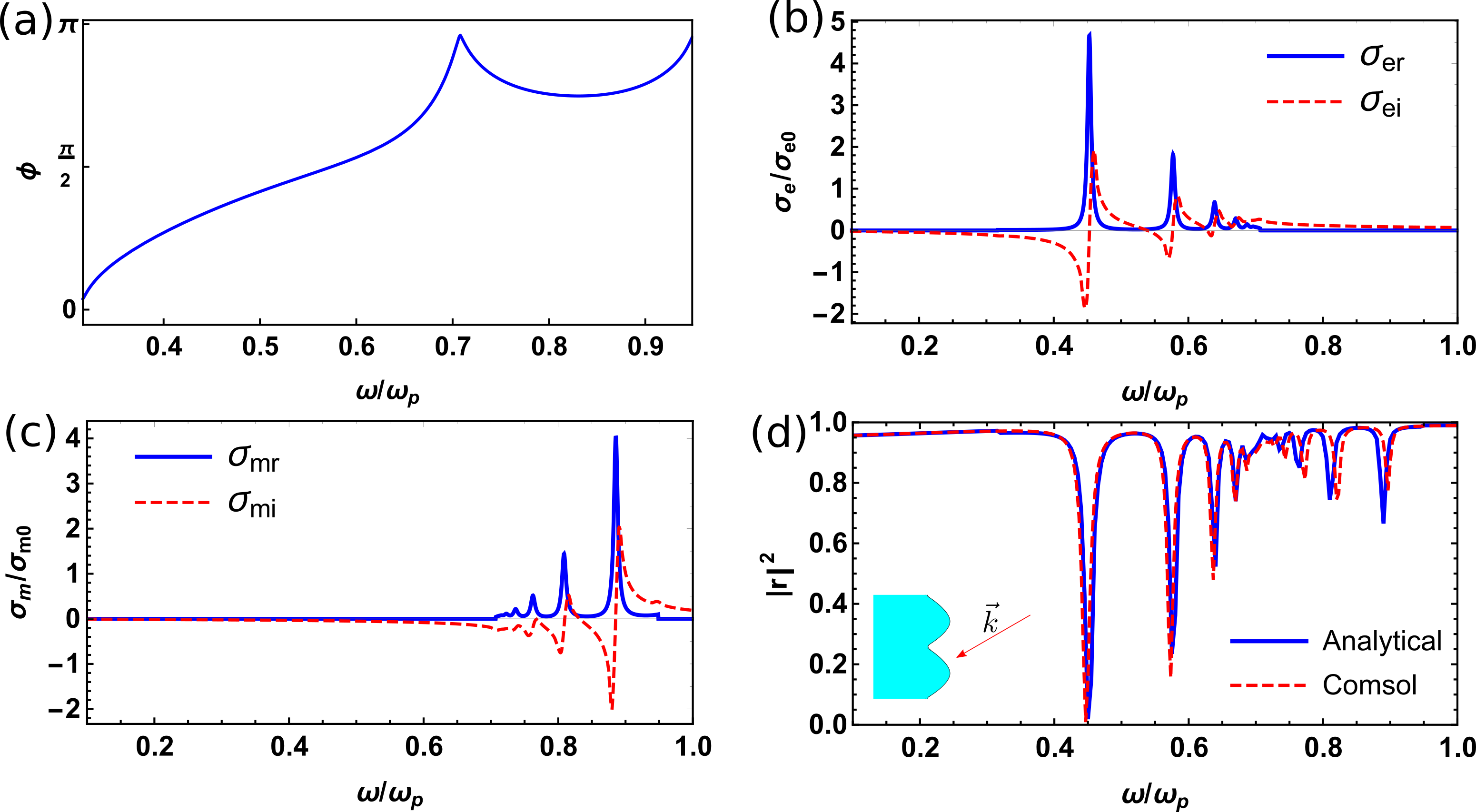}
\centering
\caption{Spectrum of a blunt singular metasurface at oblique incidence, showing a discrete rather than a continuum. (a) Phase picked-up by SPPs at the terminal of the truncated cavity, of length $L=d$. (b,c) Effective electric (magnetic) surface conductivity, real (solid blue line) and imaginary (red dashed) parts. (c) Reflectivity: analytical (solid blue line) and numerical (red dashed). The metasurface parameters are the same as in Fig. \ref{continuous_spectrum_obliqueincidence}.}
\label{discrete_spectrum_obliqueincidence}
\end{figure}

We next consider the metasurface with blunt singularities under oblique incidence. In this case we first need to calculate the phase picked-up by the SPP modes on reflection at the truncated end in the slab frame. Since both the lower and upper bands are now excited, we calculate $\phi$ not only for the anti-symmetric mode (the same as for normal incidence) but also for the symmetric mode. The SPP phase is shown in Fig. \ref{discrete_spectrum_obliqueincidence}(a), which ranges from the lower cut-off frequency $\omega_{c1}$ to higher cut-off frequency $\omega_{c2}$, and $\omega_{sp}$ clearly separates the two bands. Panels (b) and (c) present the calculated effective electric and magnetic conductivities, respectively, where the quantized modes in the truncated slab yield a discrete spectrum. Finally, we obtain the reflectivity of the metasurface which is shown in panel (d) as a blue solid line, together with results from full wave simulations (red dashed line). 
 
\subsection{\label{sec:level2}Asymmetric metasurface}
 
\begin{figure}
\includegraphics[width=0.5\columnwidth]{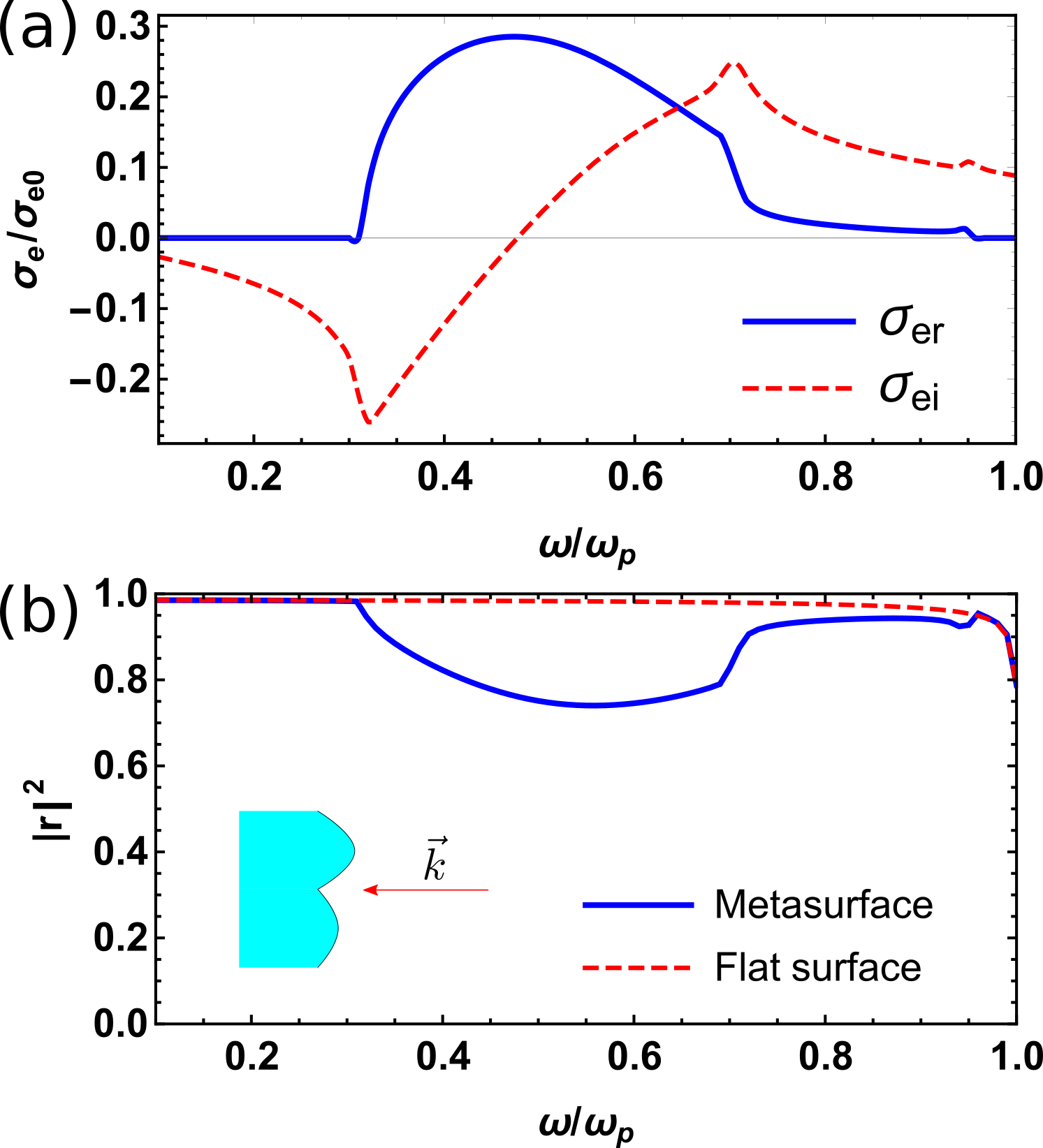}
\centering
\caption{Asymmetric singular metasurface at normal incidence. (a) Effective electric surface conductivity: real (solid blue line) and imaginary (dashed red) parts. (b) Reflectivity of the metasurface (blue solid line) and of a flat surface with the same permittivity (red dashed). The parameters used are $T=10nm$, $d_3=0.9d$, $d_1=\alpha (d-d_3)$ and $d_2=(1-\alpha)(d-d_3)$, where the asymmetry factor $\alpha$ is chosen as $0.7$.}
\label{continuous_spectrum_AsymmetricMetasurface}
\end{figure}

We now turn to discussing metasurfaces which are not symmetric with respect to $y=0$, as the one sketched in Fig. \ref{continuous_spectrum_AsymmetricMetasurface}. This asymmetric metasurface can be generated from the slab array by choosing $d_1 \neq d_2$ and, in contrast to the symmetric metasurface, it supports modes that cannot be classified as symmetric and anti-symmetric. In order to treat this case, we proceed in exactly the same way as for the symmetric metasurface and derive the mode coefficients given in Appendix E. The calculated conductivity and reflectivity spectra for a metasurface of the same period as considered previously are shown in Fig. \ref{continuous_spectrum_AsymmetricMetasurface}. Under normal incidence, the asymmetric metasurface can be modelled with an electric conductivity [see panel (a)], while $\sigma_{mr}=0$. However, as a difference with the symmetric metasurface, in this case $\sigma_{er}$ is non-zero both for the lower and upper bands. As a consequence, the calculated reflectivity spectrum (b) shows two continuous bands below and above the surface plasmon frequency, corresponding to the excitation of both the lower and upper bands. It is interesting to note that the reflectivity of the antisymmetric metasurface below $\omega_{sp}$ is very similar to that of the symmetric metasurface, specially for low frequencies. The reason for this is that the reflectivity is mainly determined by the singularity, and the groove angle is the same in both cases. 
 
\begin{figure}
\includegraphics[width=0.5\columnwidth]{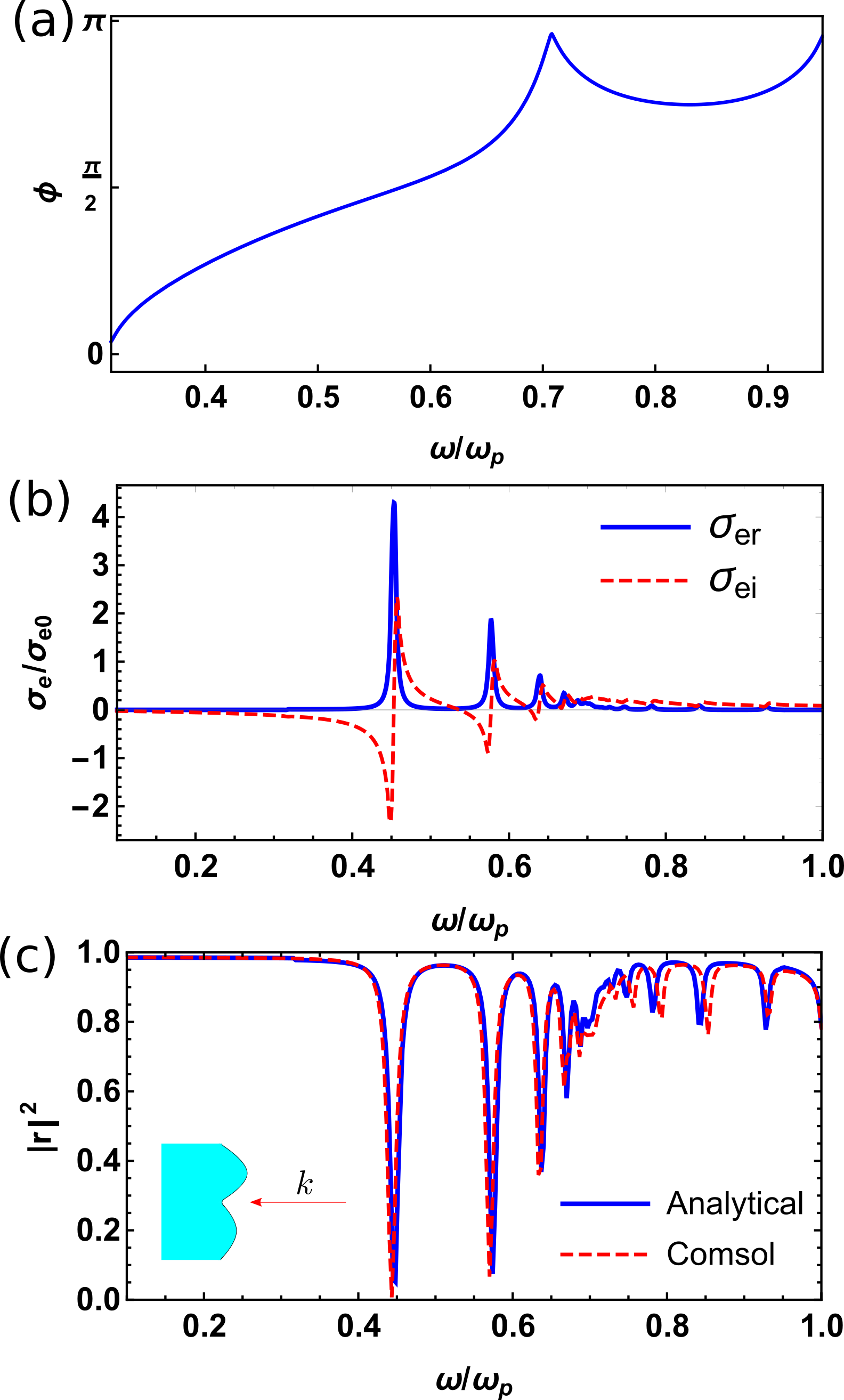}
\centering
\caption{Spectrum of a blunt asymmetric singular metasurface at normal incidence. (a) Acquired phase by the SPP mode at the terminal of the truncated cavity. (b) Effective surface conductivity: real (solid blue line) and imaginary (dashed red) parts. (c) Reflectivity: analytical (solid blue line) vs numerical (dashed red). The metasurface parameters are the same as in Fig. \ref{continuous_spectrum_AsymmetricMetasurface} and the truncation length is $L=d$. }
\label{discrete_spectrum_AsymmetricMetasurface}
\end{figure}

Finally, we also present results for an asymmetric metasurface with blunt singularities in Fig. \ref{discrete_spectrum_AsymmetricMetasurface}. Again, when the singularities are not perfect the continuous spectrum of the singular metasurface turns into a discrete spectrum. As before, we first calculate the phase acquired by SPPs in the truncated array in the slab frame [given in panel (a)], and from $\phi$ we calculate the effective electrical surface conductivity (b), which has a non-zero real part both for the lower and upper bands in this case. With this we obtain the reflectivity, which we plot in panel (c) as a solid blue line. The good agreement with results from full wave simulations, shown as a red dashed line, confirms our analytical modelling.

\section{Conclusions}
In this paper we have presented an analytical theory to study the optical response of singular plasmonic metasurfaces. By means of transformation optics, we have shown that subwavelength metasurfaces with singular grooves (or wedges) are spectrally equivalent to a more symmetric system, a simple periodic array of metal slabs. This has allowed us to obtain analytical expressions for the dispersion relation of SPP modes in the metasurface, as well as for the absorption cross section. Then, by introducing a flat surface model we have derived effective surface conductivities to model energy dissipation by the SPPs, which has enabled us to derive analytical expressions for the metasurface reflectivity. We have shown how singular plasmonic metasurfaces have continuous spectra, with a reduced reflectivity over a broad range of frequencies. On the other hand, realistic metasurfaces will not have perfect singularities, first due to the difficulties in nanofabricating sharp angles, and ultimately to non-locality. We have shown how blunt singularities have a striking effect in the spectrum of metasurfaces. Since blunt singular metasurfaces map to truncated slab arrays, the SPP modes are quantized resulting in a discrete set of peaks in the spectrum. We have also discussed how under normal incidence only the anti-symmetric band is excited, which lies below the surface plasmon frequency for the case of grooves (above for wedges). Breaking the symmetry with an incident wave at an oblique angle allows for the excitation of the symmetric band above the surface plasmon frequency (below for wedges). Finally, we have also discussed how both bands are excited at normal incidence in asymmetric metasurfaces. 
 
For metasurfaces with blunt singularities we have been able to compare our analytical results with full wave electrodynamic simulations, showing that they are in excellent agreement. At the same time, our analytical treatment gives a great physical insight in terms of singularities and allows for an easy optimization of the metasurface shape for a given purpose.

\section{Acknowledgements}
We are grateful for fruitful discussions with Yu Luo. This work was supported by the Gordon and Betty Moore Foundation. F.Y. acknowledges the Lee Family Scholarship for financial support. P.A.H. acknowledges funding from a Marie Sklodowska-Curie Fellowship.

\appendix
 
\section{\label{sec:appsource}Source fields in the transformed space}
Here we present the derivation of the incident, reflected and transmitted waves in the slab frame. Starting from plane waves in the metasurface frame, Eq. \ref{eq:source}, we apply the transformation equation (\ref{transformation_eq}) and write the source fields in the slab geometry. We assume that the metasurface is subwavelength and that the spatial region of interest satisfies $T \ll |x_4| \ll \lambda$. Under this assumption, the transformation reduces to 
\begin{equation}
\begin{split}
z_4 & \approx -\frac{T}{2\pi} \mathrm{ln} \bigg( \frac{\pi}{d}z_1 \bigg)
\end{split}
\end{equation}
for the incident and reflected waves on the right-hand side of the singular surface ($T \ll x_4 \ll \lambda$). For transmitted wave on the left side of singular surface we have $T \ll -x_4 \ll \lambda$ and,
\begin{equation}
\begin{split}
z_4 & \approx \frac{T}{2\pi} \mathrm{ln} \bigg( \frac{\pi}{d}(z_1 \pm i \frac{d}{2}) \bigg).
\end{split}
\end{equation}
 
Then, the incident wave can be written as
\begin{equation}
\begin{split}
H_z^{inc} &= H_0 e^{-i k_{0x} x_4 + i k_{0y} y_4}\\
&= H_0 e^{-i k_{0x} \frac{z_4+z_4^*}{2} + i k_{0y} \frac{z_4-z_4^*}{2 i}}\\
&= H_0 e^{\frac{-i k_{0x} + k_{0y}}{2} z_4 + \frac{-i k_{0x} - k_{0y}}{2} z_4^*}\\
&\approx H_0 e^{- \frac{-i k_{0x} + k_{0y}}{2} \frac{T}{2\pi} \mathrm{ln} \big( \frac{\pi}{d}z_1 \big)} \times e^{- \frac{-i k_{0x} - k_{0y}}{2} \frac{T}{2\pi} \mathrm{ln} \big( \frac{\pi}{d}z_1^* \big)}\\
&\approx H_0 \bigg(1 + \frac{i k_{0x} - k_{0y}}{2} \frac{T}{2\pi} \mathrm{ln} \big( \frac{\pi}{d}z_1 \big) \bigg) \times \bigg(1 + \frac{i k_{0x} + k_{0y}}{2} \frac{T}{2\pi} \mathrm{ln} \big( \frac{\pi}{d}z_1^* \big) \bigg) \\
&\approx H_0 \bigg(1 + \frac{i k_{0x} - k_{0y}}{2} \frac{T}{2\pi} \mathrm{ln} \big( \frac{\pi}{d}z_1 \big) + \frac{i k_{0x} + k_{0y}}{2} \frac{T}{2\pi} \mathrm{ln} \big( \frac{\pi}{d}z_1^* \big) \bigg) \\
&= H_0 \bigg(1 + i\frac{k_{0x}T}{4\pi} \mathrm{ln} \big( (\frac{\pi}{d})^2 |z_1|^2 \big) + \frac{k_{0y}T}{4\pi} \mathrm{ln} \big( \frac{z_1^*}{z_1} \big) \bigg) \\
&= H_0 \bigg(1 + i\frac{k_{0x}T}{2\pi} \mathrm{ln} \big( \frac{\pi}{d} \big) + i\frac{k_{0x}T}{4\pi} \mathrm{ln} \big(|z_1|^2 \big) + \frac{k_{0y}T}{4\pi} \mathrm{ln} \big( \frac{z_1^*}{z_1} \big) \bigg)\\
&= H_0 \bigg(1 + i\frac{k_{0x}T}{2\pi} \mathrm{ln} \big( \frac{\pi}{d} \big) - i\frac{k_{0x}T}{4\pi} \int\limits_{-\infty}^{\infty} \frac{e^{-|k_x||y_1|}}{|k_x|} e^{i k_x x_1} d k_x + \frac{k_{0y}T}{4\pi} \int\limits_{-\infty}^{\infty} \frac{e^{-|k_x||y_1|}}{\text{sgn}(y_1)k_x} e^{i k_x x_1} d k_x \bigg)\\
&= H_0 \bigg(1+i \frac{k_{0x} T}{2\pi} \mathrm{ln} \big( \frac{\pi}{d} \big) \bigg) +  \int\limits_{-\infty}^{\infty} a_a \frac{e^{-|k_x||y_1|}}{|k_x|} e^{i k_x x_1} d k_x +  \int\limits_{-\infty}^{\infty} a_s \frac{e^{-|k_x||y_1|}}{\text{sgn}(y_1)k_x} e^{i k_x x_1} d k_x,
\end{split}   
\end{equation}
where we have used
\begin{equation}
\begin{split}
\mathrm{ln} \big( |z_1|^2 \big) = -\int\limits_{-\infty}^{\infty} \frac{e^{-|k_x||y_1|}}{|k_x|} e^{i k_x x_1} d k_x
\end{split}   
\end{equation}
and
\begin{equation}
\begin{split}
\mathrm{ln} \big( \frac{z_1^*}{z_1} \big) = \int\limits_{-\infty}^{\infty} \frac{e^{-|k_x||y_1|}}{\text{sgn}(y_1)k_x} e^{i k_x x_1} d k_x.
\end{split}   
\end{equation}
 
The reflected wave can be easily obtained by replacing $k_{0x}$ with $-k_{0x}$,
\begin{equation}
\begin{split}
H_z^{ref} &= r H_0 e^{i k_{0x} x_4 + i k_{0y} y_4}\\
&= r H_0 \bigg(1 - i\frac{k_{0x}T}{2\pi} \mathrm{ln} \big( \frac{\pi}{d} \big) + i\frac{k_{0x}T}{4\pi} \int\limits_{-\infty}^{\infty} \frac{e^{-|k_x||y_1|}}{|k_x|} e^{i k_x x_1} d k_x + \frac{k_{0y}T}{4\pi} \int\limits_{-\infty}^{\infty} \frac{e^{-|k_x||y_1|}}{\text{sgn}(y_1)k_x} e^{i k_x x_1} d k_x \bigg)\\
&= r H_0 \bigg(1-i \frac{k_{0x} T}{2\pi} \mathrm{ln} \big( \frac{\pi}{d} \big) \bigg) -  \int\limits_{-\infty}^{\infty} r a_a \frac{e^{-|k_x||y_1|}}{|k_x|} e^{i k_x x_1} d k_x +  \int\limits_{-\infty}^{\infty} r a_s \frac{e^{-|k_x||y_1|}}{\text{sgn}(y_1)k_x} e^{i k_x x_1} d k_x
\end{split}   
\end{equation}
 
For the transmitted wave, $k_{0x}^{'} = \sqrt{\varepsilon k_0^2 - k_{0y}^2}$, and we have 
\begin{equation}
\begin{split}
H_z^{tra} &= t H_0 e^{-i k_{0x}^{'}  x_4 + i k_{0y} y_4}\\
&= t H_0 e^{-i k_{0x}^{'} \frac{z_4+z_4^*}{2} + i k_{0y} \frac{z_4-z_4^*}{2 i}}\\
&= t H_0 e^{\frac{-i k_{0x}^{'} + k_{0y}}{2} z_4 + \frac{-i k_{0x}^{'} - k_{0y}}{2} z_4^*}\\
&\approx t H_0 e^{\frac{-i k_{0x}^{'} + k_{0y}}{2} \frac{T}{2\pi} \mathrm{ln} \big( \frac{\pi}{d}(z_1 + i \frac{d}{2}) \big)} \times e^{ \frac{-i k_{0x}^{'} - k_{0y}}{2} \frac{T}{2\pi} \mathrm{ln} \big( \frac{\pi}{d}(z_1 + i \frac{d}{2})^* \big)}\\
&\approx t H_0 \bigg(1 - \frac{i k_{0x}^{'} - k_{0y}}{2} \frac{T}{2\pi} \mathrm{ln} \big( \frac{\pi}{d}(z_1+i\frac{d}{2}) \big) \bigg) \times \bigg(1 - \frac{i k_{0x}^{'} + k_{0y}}{2} \frac{T}{2\pi} \mathrm{ln} \big( \frac{\pi}{d}(z_1+i\frac{d}{2})^* \big) \bigg) \\
&\approx t H_0 \bigg(1 - \frac{i k_{0x}^{'} - k_{0y}}{2} \frac{T}{2\pi} \mathrm{ln} \big( \frac{\pi}{d}(z_1+i\frac{d}{2}) \big) - \frac{i k_{0x}^{'} + k_{0y}}{2} \frac{T}{2\pi} \mathrm{ln} \big( \frac{\pi}{d}(z_1+i\frac{d}{2})^* \big) \bigg) \\
&= t H_0 \bigg(1 - i\frac{k_{0x}^{'}T}{4\pi} \mathrm{ln} \big( (\frac{\pi}{d})^2 |z_1+i\frac{d}{2}|^2 \big) - \frac{k_{0y}T}{4\pi} \mathrm{ln} \big( \frac{(z_1+i\frac{d}{2})^*}{(z_1+i\frac{d}{2})} \big) \bigg) \\
&= t H_0 \bigg(1 - i\frac{k_{0x}^{'}T}{2\pi} \mathrm{ln} \big( \frac{\pi}{d} \big) - i\frac{k_{0x}^{'}T}{4\pi} \mathrm{ln} \big( |z_1+i\frac{d}{2}|^2 \big) - \frac{k_{0y}T}{4\pi} \mathrm{ln} \big( \frac{(z_1+i\frac{d}{2})^*}{(z_1+i\frac{d}{2})} \big) \bigg) \\
&= t H_0 \bigg(1 - i\frac{k_{0x}^{'}T}{2\pi} \mathrm{ln} \big( \frac{\pi}{d} \big) + i\frac{k_{0x}^{'}T}{4\pi} \int\limits_{-\infty}^{\infty} \frac{e^{-|k_x||y_1+\frac{d}{2}|}}{|k_x|} e^{i k_x x_1} d k_x - \frac{k_{0y}T}{4\pi} \int\limits_{-\infty}^{\infty} \frac{e^{-|k_x||y_1+\frac{d}{2}|}}{\text{sgn}(y_1+\frac{d}{2})k_x} e^{i k_x x_1} d k_x \bigg)\\
&= t H_0 \bigg(1 - i \frac{k_{0x}^{'} T}{2\pi} \mathrm{ln} \big( \frac{\pi}{d} \big) \bigg) - \int\limits_{-\infty}^{\infty} t \frac{k_{0x}^{'}}{k_{0x}}  a_a \frac{e^{-|k_x||y_1|}}{|k_x|} e^{i k_x x_1} d k_x  -  \int\limits_{-\infty}^{\infty} t a_s \frac{e^{-|k_x||y_1+\frac{d}{2}|}}{\text{sgn}(y_1+\frac{d}{2})k_x} e^{i k_x x_1} d k_x
\end{split}   
\end{equation}
 
\section{\label{sec:appamplitude}Calculation of the SPP mode coefficients}
In order to calculate the mode coefficients of the excited SPPs, we apply the boundary conditions in the slab frame. Hence, we match the tangential field components, $H_z$ and $E_x$, at the boundaries $y=d_1$, $y=-d_2$, $y=-(d_2+d_3)$ \cite{luo2010surface}. In the matrix form, the equations system reads as,
\begin{equation}
\begin{split}
    &\left( {\begin{array}{*{20}{l}}
    e^{|k_x|d_2} & e^{-|k_x|d_2} & -e^{|k_x|d_2} & -e^{-|k_x|d_2} \\
    e^{-|k_x|d_1} & e^{|k_x|d_1} & -e^{|k_x|(d_2+d_3)} & -e^{-|k_x|(d_2+d_3)} \\
    |k_x|e^{|k_x|d_2} & -|k_x|e^{-|k_x|d_2} & -\frac{|k_x|e^{|k_x|d_2}}{\varepsilon} & \frac{|k_x|e^{-|k_x|d_2}}{\varepsilon} \\
    |k_x|e^{-|k_x|d_1} & -|k_x|e^{|k_x|d_1} & -\frac{|k_x|e^{|k_x|(d_2+d_3)}}{\varepsilon} & \frac{|k_x|e^{-|k_x|(d_2+d_3)}}{\varepsilon}
    \end{array}} \right)
    \left( {\begin{array}{*{20}{l}}
    b_+ \\ b_- \\ c_+ \\ c_-
    \end{array}} \right)
    \\
    &=
    \left( {\begin{array}{*{10}{c}}
    -\frac{e^{-|k_x|d_2}}{|k_x|}\\ -\frac{e^{-|k_x|d_1}}{|k_x|} \\ e^{-|k_x|d_2} \\ - e^{-|k_x|d_1}
    \end{array}} \right) a_a (1-r)
    +
    \left( {\begin{array}{*{10}{c}}
    -\frac{e^{-|k_x|(-d_2+\frac{d}{2})}}{|k_x|}\\ -\frac{e^{-|k_x|(d_2+d_3-\frac{d}{2})}}{|k_x|} \\ -\frac{e^{-|k_x|(-d_2+\frac{d}{2})}}{\varepsilon} \\ \frac{e^{-|k_x|(d_2+d_3-\frac{d}{2})}}{\varepsilon}
    \end{array}} \right) a_a \frac{k_{0x}^{'}}{k_{0x}} t
    \\
    &+
    \left( {\begin{array}{*{10}{c}}
    \frac{e^{-|k_x|d_2}}{k_x}\\ -\frac{e^{-|k_x|d_1}}{k_x} \\ - \text{sgn}(k_x) e^{-|k_x|d_2} \\ - \text{sgn}(k_x) e^{-|k_x|d_1}
    \end{array}} \right) a_s (1+r)
    +
    \left( {\begin{array}{*{10}{c}}
    -\frac{e^{-|k_x|(-d_2+\frac{d}{2})}}{k_x}\\ \frac{e^{-|k_x|(d_2+d_3-\frac{d}{2})}}{k_x} \\ 
    - \text{sgn}(k_x) \frac{e^{-|k_x|(-d_2+\frac{d}{2})}}{\varepsilon} \\ 
    - \text{sgn}(k_x) \frac{e^{-|k_x|(d_2+d_3-\frac{d}{2})}}{\varepsilon}
    \end{array}} \right) a_s t
\end{split}
\end{equation}
where the Bloch wave condition has been used because of the periodicity of the slab array. Since the system of equations is linear, the response functions $b$ and $c$ can be decomposed into anti-symmetric and symmetric components. For the anti-symmetric excitation ($a_a \neq 0$, $a_s = 0$), we have
\begin{equation}
\begin{split}
        b_{a+} =& \frac{(\varepsilon -1)e^{|k_x| d_3}+\varepsilon +1}{|k_x| ((\varepsilon-1)(e^{|k_x|(d_1+d_2)} - e^{|k_x|d_3}) + (\varepsilon+1)(e^{|k_x|d}-1))} a_a (1-r) \\
        & -\frac{2 e^{\frac{1}{2}|k_x| d}}{|k_x| ((\varepsilon-1)(e^{|k_x|(d_1+d_2)} - e^{|k_x|d_3}) + (\varepsilon+1)(e^{|k_x|d}-1))} a_a \frac{k_{0x}^{'}}{k_{0x}} t \\
        b_{a-} =& \frac{(\varepsilon -1)e^{|k_x| d_3}+\varepsilon +1}{|k_x| ((\varepsilon-1)(e^{|k_x|(d_1+d_2)} - e^{|k_x|d_3}) + (\varepsilon+1)(e^{|k_x|d}-1))} a_a (1-r) \\
        & -\frac{2 e^{\frac{1}{2}|k_x| d}}{|k_x| ((\varepsilon-1)(e^{|k_x|(d_1+d_2)} - e^{|k_x|d_3}) + (\varepsilon+1)(e^{|k_x|d}-1))} a_a \frac{k_{0x}^{'}}{k_{0x}} t
\end{split}
\end{equation}
\begin{equation}
\begin{split}
        c_{a+} =& \frac{2\varepsilon}{|k_x| ((\varepsilon-1)(e^{|k_x|(d_1+d_2)} - e^{|k_x|d_3}) + (\varepsilon+1)(e^{|k_x|d}-1))} a_a (1-r) \\
        & + \frac{e^{-\frac{1}{2}|k_x| d}((\varepsilon-1)e^{2|k_x| d_2}-(\varepsilon+1))}{|k_x| ((\varepsilon-1)(e^{|k_x|(d_1+d_2)} - e^{|k_x|d_3}) + (\varepsilon+1)(e^{|k_x|d}-1))} a_a \frac{k_{0x}^{'}}{k_{0x}} t \\
        c_{a-} =& \frac{2 \varepsilon  e^{|k_x| d}}{|k_x| ((\varepsilon-1)(e^{|k_x|(d_1+d_2)} - e^{|k_x|d_3}) + (\varepsilon+1)(e^{|k_x|d}-1))} a_a (1-r) \\
        & + \frac{e^{\frac{1}{2}|k_x| d}((\varepsilon-1)e^{2|k_x| d_2}-(\varepsilon+1))}{|k_x| ((\varepsilon-1)(e^{|k_x|(d_1+d_2)} - e^{|k_x|d_3}) + (\varepsilon+1)(e^{|k_x|d}-1))} a_a \frac{k_{0x}^{'}}{k_{0x}} t
\end{split}
\end{equation}
 
On the other hand, the symmetric excitation ($a_a = 0$, $a_s \neq 0$) gives
\begin{equation}
\begin{split}
        b_{s+} =& \frac{(\varepsilon-1)e^{|k_x|d_3} - (\varepsilon+1)}{|k_x| ((\varepsilon-1)(e^{|k_x|(d_1+d_2)} - e^{|k_x|d_3}) - (\varepsilon+1)(e^{|k_x|d}-1))} \text{sgn}(k_x) a_s (1+r) \\
        & + \frac{2 e^{\frac{1}{2}|k_x|d}}{|k_x| ((\varepsilon-1)(e^{|k_x|(d_1+d_2)} - e^{|k_x|d_3}) - (\varepsilon+1)(e^{|k_x|d}-1))}\text{sgn}(k_x) a_s t \\
        b_{s-} =& -\frac{(\varepsilon-1)e^{|k_x|d_3} - (\varepsilon+1)}{|k_x| ((\varepsilon-1)(e^{|k_x|(d_1+d_2)} - e^{|k_x|d_3}) - (\varepsilon+1)(e^{|k_x|d}-1))} \text{sgn}(k_x) a_s (1+r) \\
        & -\frac{2 e^{\frac{1}{2}|k_x|d}}{|k_x| ((\varepsilon-1)(e^{|k_x|(d_1+d_2)} - e^{|k_x|d_3}) - (\varepsilon+1)(e^{|k_x|d}-1))}\text{sgn}(k_x) a_s t
\end{split}
\end{equation}
\begin{equation}
\begin{split}  
        c_{s+} =& -\frac{2\varepsilon}{|k_x| ((\varepsilon-1)(e^{|k_x|(d_1+d_2)} - e^{|k_x|d_3}) - (\varepsilon+1)(e^{|k_x|d}-1))} \text{sgn}(k_x) a_s (1+r) \\
        & +\frac{e^{-\frac{1}{2}|k_x|d}((\varepsilon-1)e^{2 d_2 |k_x|}+ (\varepsilon+1))}{|k_x| ((\varepsilon-1)(e^{|k_x|(d_1+d_2)} - e^{|k_x|d_3}) - (\varepsilon+1)(e^{|k_x|d}-1))} \text{sgn}(k_x) a_s t \\
        c_{s-} =& \frac{2\varepsilon e^{|k_x|d}}{|k_x| ((\varepsilon-1)(e^{|k_x|(d_1+d_2)} - e^{|k_x|d_3}) - (\varepsilon+1)(e^{|k_x|d}-1))} \text{sgn}(k_x) a_s (1+r) \\
        & -\frac{e^{\frac{1}{2}|k_x|d}((\varepsilon-1)e^{2 d_2 |k_x|}+ (\varepsilon+1))}{|k_x| ((\varepsilon-1)(e^{|k_x|(d_1+d_2)} - e^{|k_x|d_3}) - (\varepsilon+1)(e^{|k_x|d}-1))} \text{sgn}(k_x) a_s  t
\end{split}
\end{equation}
 
Then, the field amplitude in the real space is obtained through a Fourier transformation of the field in k-space, as discussed in the main text. Here we give the mode coefficients of the fields in real space. For the anti-symmetric mode, we have,
\begin{equation}
\begin{split}
\Gamma_{a+}  =& \frac{(\varepsilon -1)e^{\sqrt{k_{px}^2} d_3}+\varepsilon +1}{k_{px} ((\varepsilon-1)((d_1+d_2)e^{\sqrt{k_{px}^2}(d_1+d_2)} - d_3 e^{\sqrt{k_{px}^2}d_3}) + (\varepsilon+1)d e^{\sqrt{k_{px}^2}d})}(1-r) \\
& -\frac{2 e^{\frac{1}{2}\sqrt{k_{px}^2} d}}{k_{px} ((\varepsilon-1)((d_1+d_2)e^{\sqrt{k_{px}^2}(d_1+d_2)} - d_3 e^{\sqrt{k_{px}^2}d_3}) + (\varepsilon+1)d e^{\sqrt{k_{px}^2}d})} \frac{k_{0x}^{'}}{k_{0x}} t  \\
\Gamma_{a-}  =& \frac{(\varepsilon -1)e^{\sqrt{k_{px}^2} d_3}+\varepsilon +1}{k_{px} ((\varepsilon-1)((d_1+d_2)e^{\sqrt{k_{px}^2}(d_1+d_2)} - d_3 e^{\sqrt{k_{px}^2}d_3}) + (\varepsilon+1)d e^{\sqrt{k_{px}^2}d})}(1-r) \\
& -\frac{2 e^{\frac{1}{2}\sqrt{k_{px}^2} d}}{k_{px} ((\varepsilon-1)((d_1+d_2)e^{\sqrt{k_{px}^2}(d_1+d_2)} - d_3 e^{\sqrt{k_{px}^2}d_3}) + (\varepsilon+1)d e^{\sqrt{k_{px}^2}d})} \frac{k_{0x}^{'}}{k_{0x}} t
\end{split} 
\end{equation}
\begin{equation}
\begin{split}
\Lambda_{a+}  =& \frac{2\varepsilon}{k_{px} ((\varepsilon-1)((d_1+d_2)e^{\sqrt{k_{px}^2}(d_1+d_2)} - d_3 e^{\sqrt{k_{px}^2}d_3}) + (\varepsilon+1)d e^{\sqrt{k_{px}^2}d})}(1-r) \\
& + \frac{e^{-\frac{1}{2}\sqrt{k_{px}^2} d}((\varepsilon-1)e^{2\sqrt{k_{px}^2} d_2}-(\varepsilon+1))}{k_{px} ((\varepsilon-1)((d_1+d_2)e^{\sqrt{k_{px}^2}(d_1+d_2)} - d_3 e^{\sqrt{k_{px}^2}d_3}) + (\varepsilon+1)d e^{\sqrt{k_{px}^2}d})} \frac{k_{0x}^{'}}{k_{0x}} t \\
\Lambda_{a-}  =& \frac{2 \varepsilon  e^{\sqrt{k_{px}^2}d}}{k_{px} ((\varepsilon-1)((d_1+d_2)e^{\sqrt{k_{px}^2}(d_1+d_2)} - d_3 e^{\sqrt{k_{px}^2}d_3}) + (\varepsilon+1)d e^{\sqrt{k_{px}^2}d})}(1-r)\\
& + \frac{e^{\frac{1}{2}\sqrt{k_{px}^2} d}((\varepsilon-1)e^{2\sqrt{k_{px}^2} d_2}-(\varepsilon+1))}{k_{px} ((\varepsilon-1)((d_1+d_2)e^{\sqrt{k_{px}^2}(d_1+d_2)} - d_3 e^{\sqrt{k_{px}^2}d_3}) + (\varepsilon+1)d e^{\sqrt{k_{px}^2}d})} \frac{k_{0x}^{'}}{k_{0x}} t
\end{split} 
\end{equation}
 
While for the symmetric mode, 
\begin{equation}
\begin{split}
        \Gamma_{s+} =& \frac{(\varepsilon-1)e^{\sqrt{k_{px}^2}d_3} - (\varepsilon+1)}{k_{px} ((\varepsilon-1)((d_1+d_2)e^{\sqrt{k_{px}^2}(d_1+d_2)} - d_3 e^{\sqrt{k_{px}^2}d_3}) - (\varepsilon+1)d e^{\sqrt{k_{px}^2}d})} \text{sgn}(k_{px})\text{sgn}(x) (1+r) \\
        & + \frac{2 e^{\frac{1}{2}\sqrt{k_{px}^2}d}}{k_{px} ((\varepsilon-1)((d_1+d_2)e^{\sqrt{k_{px}^2}(d_1+d_2)} - d_3 e^{\sqrt{k_{px}^2}d_3}) - (\varepsilon+1)d e^{\sqrt{k_{px}^2}d})} \text{sgn}(k_{px})\text{sgn}(x) t \\
        \Gamma_{s-} =& -\frac{(\varepsilon-1)e^{\sqrt{k_{px}^2}d_3} - (\varepsilon+1)}{k_{px} ((\varepsilon-1)((d_1+d_2)e^{\sqrt{k_{px}^2}(d_1+d_2)} - d_3 e^{\sqrt{k_{px}^2}d_3}) - (\varepsilon+1)d e^{\sqrt{k_{px}^2}d})} \text{sgn}(k_{px})\text{sgn}(x) (1+r) \\
        & -\frac{2 e^{\frac{1}{2}\sqrt{k_{px}^2}d}}{k_{px} ((\varepsilon-1)((d_1+d_2)e^{\sqrt{k_{px}^2}(d_1+d_2)} - d_3 e^{\sqrt{k_{px}^2}d_3}) - (\varepsilon+1)d e^{\sqrt{k_{px}^2}d})} \text{sgn}(k_{px})\text{sgn}(x) t
\end{split} 
\end{equation}
\begin{equation}
\begin{split}
        \Lambda_{s+} =& -\frac{2\varepsilon}{k_{px} ((\varepsilon-1)((d_1+d_2)e^{\sqrt{k_{px}^2}(d_1+d_2)} - d_3 e^{\sqrt{k_{px}^2}d_3}) - (\varepsilon+1)d e^{\sqrt{k_{px}^2}d})} \text{sgn}(k_{px})\text{sgn}(x) (1+r) \\
        & +\frac{e^{-\frac{1}{2}\sqrt{k_{px}^2}d}((\varepsilon-1)e^{2 d_2 \sqrt{k_{px}^2}}+ (\varepsilon+1))}{k_{px} ((\varepsilon-1)((d_1+d_2)e^{\sqrt{k_{px}^2}(d_1+d_2)} - d_3 e^{\sqrt{k_{px}^2}d_3}) - (\varepsilon+1)d e^{\sqrt{k_{px}^2}d})} \text{sgn}(k_{px})\text{sgn}(x) t \\
        \Lambda_{s-} =& \frac{2\varepsilon e^{\sqrt{k_{px}^2}d}}{k_{px} ((\varepsilon-1)((d_1+d_2)e^{\sqrt{k_{px}^2}(d_1+d_2)} - d_3 e^{\sqrt{k_{px}^2}d_3}) - (\varepsilon+1)d e^{\sqrt{k_{px}^2}d})} \text{sgn}(k_{px})\text{sgn}(x) (1+r) \\
        & -\frac{e^{\frac{1}{2}\sqrt{k_{px}^2}d}((\varepsilon-1)e^{2 d_2 \sqrt{k_{px}^2}}+ (\varepsilon+1))}{k_{px} ((\varepsilon-1)((d_1+d_2)e^{\sqrt{k_{px}^2}(d_1+d_2)} - d_3 e^{\sqrt{k_{px}^2}d_3}) - (\varepsilon+1)d e^{\sqrt{k_{px}^2}d})} \text{sgn}(k_{px})\text{sgn}(x)  t
\end{split}
\end{equation}

\section{\label{sec:appnormalizedamplitude}Normalized mode amplitudes}
In the flat surface model introduced in the main text, we have that the mode coefficients ($\Gamma_{(a,s)\pm}$, $\Lambda_{(a,s)\pm}$) are proportional to $1-r$ for the anti-symmetric mode and to $1+r$ to the symmetric mode, such that normalized coefficients can be defined which do not depend on $r$ or $t$, denoted as $\Gamma_{(a,s)\pm}^{'}$, $\Lambda_{(a,s)\pm}^{'}$. 
 
For the anti-symmetric mode, we have
\begin{equation}
\begin{split}
\Gamma_{a+}^{'}  =& \frac{(\varepsilon -1)e^{\sqrt{k_{px}^2} d_3}+\varepsilon +1}{k_{px} ((\varepsilon-1)((d_1+d_2)e^{\sqrt{k_{px}^2}(d_1+d_2)} - d_3 e^{\sqrt{k_{px}^2}d_3}) + (\varepsilon+1)d e^{\sqrt{k_{px}^2}d})} \\
& -\frac{2 e^{\frac{1}{2}\sqrt{k_{px}^2} d}}{k_{px} ((\varepsilon-1)((d_1+d_2)e^{\sqrt{k_{px}^2}(d_1+d_2)} - d_3 e^{\sqrt{k_{px}^2}d_3}) + (\varepsilon+1)d e^{\sqrt{k_{px}^2}d})} \varepsilon  \\
\Gamma_{a-}^{'}  =& \frac{(\varepsilon -1)e^{\sqrt{k_{px}^2} d_3}+\varepsilon +1}{k_{px} ((\varepsilon-1)((d_1+d_2)e^{\sqrt{k_{px}^2}(d_1+d_2)} - d_3 e^{\sqrt{k_{px}^2}d_3}) + (\varepsilon+1)d e^{\sqrt{k_{px}^2}d})} \\
& -\frac{2 e^{\frac{1}{2}\sqrt{k_{px}^2} d}}{k_{px} ((\varepsilon-1)((d_1+d_2)e^{\sqrt{k_{px}^2}(d_1+d_2)} - d_3 e^{\sqrt{k_{px}^2}d_3}) + (\varepsilon+1)d e^{\sqrt{k_{px}^2}d})} \varepsilon 
\end{split} 
\end{equation}
\begin{equation}
\begin{split}
\Lambda_{a+}^{'}  =& \frac{2\varepsilon}{k_{px} ((\varepsilon-1)((d_1+d_2)e^{\sqrt{k_{px}^2}(d_1+d_2)} - d_3 e^{\sqrt{k_{px}^2}d_3}) + (\varepsilon+1)d e^{\sqrt{k_{px}^2}d})}\\
& + \frac{e^{-\frac{1}{2}\sqrt{k_{px}^2} d}((\varepsilon-1)e^{2\sqrt{k_{px}^2} d_2}-(\varepsilon+1))}{k_{px} ((\varepsilon-1)((d_1+d_2)e^{\sqrt{k_{px}^2}(d_1+d_2)} - d_3 e^{\sqrt{k_{px}^2}d_3}) + (\varepsilon+1)d e^{\sqrt{k_{px}^2}d})} \varepsilon \\
\Lambda_{a-}^{'}  =& \frac{2 \varepsilon  e^{\sqrt{k_{px}^2}d}}{k_{px} ((\varepsilon-1)((d_1+d_2)e^{\sqrt{k_{px}^2}(d_1+d_2)} - d_3 e^{\sqrt{k_{px}^2}d_3}) + (\varepsilon+1)d e^{\sqrt{k_{px}^2}d})}\\
& + \frac{e^{\frac{1}{2}\sqrt{k_{px}^2} d}((\varepsilon-1)e^{2\sqrt{k_{px}^2} d_2}-(\varepsilon+1))}{k_{px} ((\varepsilon-1)((d_1+d_2)e^{\sqrt{k_{px}^2}(d_1+d_2)} - d_3 e^{\sqrt{k_{px}^2}d_3}) + (\varepsilon+1)d e^{\sqrt{k_{px}^2}d})} \varepsilon 
\end{split} 
\end{equation}
 
Similarly, for the symmetric mode, we have 
\begin{equation}
\begin{split}
        \Gamma_{s+}^{'} =& \frac{(\varepsilon-1)e^{\sqrt{k_{px}^2}d_3} - (\varepsilon+1)}{k_{px} ((\varepsilon-1)((d_1+d_2)e^{\sqrt{k_{px}^2}(d_1+d_2)} - d_3 e^{\sqrt{k_{px}^2}d_3}) - (\varepsilon+1)d e^{\sqrt{k_{px}^2}d})} \text{sgn}(k_{px})\text{sgn}(x) \\
        & + \frac{2 e^{\frac{1}{2}\sqrt{k_{px}^2}d}}{k_{px} ((\varepsilon-1)((d_1+d_2)e^{\sqrt{k_{px}^2}(d_1+d_2)} - d_3 e^{\sqrt{k_{px}^2}d_3}) - (\varepsilon+1)d e^{\sqrt{k_{px}^2}d})} \text{sgn}(k_{px})\text{sgn}(x) \\
        \Gamma_{s-}^{'} =& -\frac{(\varepsilon-1)e^{\sqrt{k_{px}^2}d_3} - (\varepsilon+1)}{k_{px} ((\varepsilon-1)((d_1+d_2)e^{\sqrt{k_{px}^2}(d_1+d_2)} - d_3 e^{\sqrt{k_{px}^2}d_3}) - (\varepsilon+1)d e^{\sqrt{k_{px}^2}d})} \text{sgn}(k_{px})\text{sgn}(x) \\
        & -\frac{2 e^{\frac{1}{2}\sqrt{k_{px}^2}d}}{k_{px} ((\varepsilon-1)((d_1+d_2)e^{\sqrt{k_{px}^2}(d_1+d_2)} - d_3 e^{\sqrt{k_{px}^2}d_3}) - (\varepsilon+1)d e^{\sqrt{k_{px}^2}d})} \text{sgn}(k_{px})\text{sgn}(x)
\end{split} 
\end{equation}
\begin{equation}
\begin{split}
        \Lambda_{s+}^{'} =& -\frac{2\varepsilon}{k_{px} ((\varepsilon-1)((d_1+d_2)e^{\sqrt{k_{px}^2}(d_1+d_2)} - d_3 e^{\sqrt{k_{px}^2}d_3}) - (\varepsilon+1)d e^{\sqrt{k_{px}^2}d})} \text{sgn}(k_{px})\text{sgn}(x) \\
        & +\frac{e^{-\frac{1}{2}\sqrt{k_{px}^2}d}((\varepsilon-1)e^{2 d_2 \sqrt{k_{px}^2}}+ (\varepsilon+1))}{k_{px} ((\varepsilon-1)((d_1+d_2)e^{\sqrt{k_{px}^2}(d_1+d_2)} - d_3 e^{\sqrt{k_{px}^2}d_3}) - (\varepsilon+1)d e^{\sqrt{k_{px}^2}d})} \text{sgn}(k_{px})\text{sgn}(x) \\
        \Lambda_{s-}^{'} =& \frac{2\varepsilon e^{\sqrt{k_{px}^2}d}}{k_{px} ((\varepsilon-1)((d_1+d_2)e^{\sqrt{k_{px}^2}(d_1+d_2)} - d_3 e^{\sqrt{k_{px}^2}d_3}) - (\varepsilon+1)d e^{\sqrt{k_{px}^2}d})} \text{sgn}(k_{px})\text{sgn}(x) \\
        & -\frac{e^{\frac{1}{2}\sqrt{k_{px}^2}d}((\varepsilon-1)e^{2 d_2 \sqrt{k_{px}^2}}+ (\varepsilon+1))}{k_{px} ((\varepsilon-1)((d_1+d_2)e^{\sqrt{k_{px}^2}(d_1+d_2)} - d_3 e^{\sqrt{k_{px}^2}d_3}) - (\varepsilon+1)d e^{\sqrt{k_{px}^2}d})} \text{sgn}(k_{px})\text{sgn}(x) 
\end{split}
\end{equation}

\section{\label{sec:appAbsblunt} Intrinsic absorption cross section for metasurfaces with blunt singularities}
Here we give expressions for the intrinsic absorption cross section of a metasurface with blunt singularities. For the anti-symmetric mode,
\begin{equation}
\begin{split}
    \sigma_{abs}^{a'} &= \frac{k_0 T}{2} \frac{|k_{px}|^2 \mathrm{Im}[\varepsilon]}{|\varepsilon|^2} \frac{1}{|1-e^{i k_{px} L + i \phi}|^2} 
    \bigg[ (\frac{|\Lambda_{a+}^{'}|^2}{-2 \mathrm{Re}[\sqrt{k_{px}^2}]} (e^{2 \mathrm{Re}[\sqrt{k_{px}^2}]d_2}-e^{2 \mathrm{Re}[\sqrt{k_{px}^2}](d_2+d_3)}) \\
    & + \frac{|\Lambda_{a-}^{'}|^2}{2 \mathrm{Re}[\sqrt{k_{px}^2}]} (e^{-2 \mathrm{Re}[\sqrt{k_{px}^2}]d_2}-e^{-2 \mathrm{Re}[\sqrt{k_{px}^2}](d_2+d_3)})) \frac{1-e^{-2\mathrm{Im}[k_{px}]L}}{\mathrm{Im}[k_{px}]} \\
    &- (\frac{\Lambda_{a+}^{'*} \Lambda_{a-}^{'}}{i 2 \mathrm{Im}[\sqrt{k_{px}^2}]} (e^{-i 2 \mathrm{Im}[\sqrt{k_{px}^2}]d_2} - e^{-i 2 \mathrm{Im}[\sqrt{k_{px}^2}](d_2+d_3)}) \\
    &+ \frac{\Lambda_{a+}^{'} \Lambda_{a-}^{'*}}{-i 2 \mathrm{Im}[\sqrt{k_{px}^2}]} (e^{i 2 \mathrm{Im}[\sqrt{k_{px}^2}]d_2} - e^{i 2 \mathrm{Im}[\sqrt{k_{px}^2}](d_2+d_3)})) \frac{2 e^{-\mathrm{Im}[k_{px}]L}}{\mathrm{Re}[k_{px}]}(\sin(\mathrm{Re}[k_{px}]L+\phi)-\sin(\phi)) \bigg]
\end{split}
\end{equation}
 
For the symmetric mode,
\begin{equation}
\begin{split}
    \sigma_{abs}^{s'} &= \frac{k_0 T}{2} \frac{|k_{px}|^2 \mathrm{Im}[\varepsilon]}{|\varepsilon|^2} \frac{1}{|1+e^{i k_{px} L + i \phi}|^2} 
    \bigg[ (\frac{|\Lambda_{s+}^{'}|^2}{-2 \mathrm{Re}[\sqrt{k_{px}^2}]} (e^{2 \mathrm{Re}[\sqrt{k_{px}^2}]d_2}-e^{2 \mathrm{Re}[\sqrt{k_{px}^2}](d_2+d_3)}) \\
    & + \frac{|\Lambda_{s-}^{'}|^2}{2 \mathrm{Re}[\sqrt{k_{px}^2}]} (e^{-2 \mathrm{Re}[\sqrt{k_{px}^2}]d_2}-e^{-2 \mathrm{Re}[\sqrt{k_{px}^2}](d_2+d_3)})) \frac{1-e^{-2\mathrm{Im}[k_{px}]L}}{\mathrm{Im}[k_{px}]} \\
    &- (\frac{\Lambda_{s+}^{'*} \Lambda_{s-}^{'}}{i 2 \mathrm{Im}[\sqrt{k_{px}^2}]} (e^{-i 2 \mathrm{Im}[\sqrt{k_{px}^2}]d_2} - e^{-i 2 \mathrm{Im}[\sqrt{k_{px}^2}](d_2+d_3)}) \\
    &+ \frac{\Lambda_{s+}^{'} \Lambda_{s-}^{'*}}{-i 2 \mathrm{Im}[\sqrt{k_{px}^2}]} (e^{i 2 \mathrm{Im}[\sqrt{k_{px}^2}]d_2} - e^{i 2 \mathrm{Im}[\sqrt{k_{px}^2}](d_2+d_3)})) \frac{2 e^{-\mathrm{Im}[k_{px}]L}}{\mathrm{Re}[k_{px}]}(\sin(\mathrm{Re}[k_{px}]L+\phi)-\sin(\phi)) \bigg]
\end{split}
\end{equation}
 
\section{\label{sec:appAsymmetric}Coefficients for the asymmetric metasurface}
Following the same procedure as the calculation for the symmetric metasurface, we obtain all the coefficients for the asymmetric metasurface ($b_{a\pm}$, $c_{a\pm}$, $\Gamma_{a\pm}$, $\Lambda_{a\pm}$, $\Lambda_{a\pm}^{'}$). They are listed as below
{\tiny
\begin{equation}
\begin{split}
        b_{a+} =& \frac{-4 \varepsilon  e^{|k_x| (d_1+d_2+d_3)}-\left(\varepsilon ^2-1\right) e^{2 |k_x| (d_1+d_3)}+\left(\varepsilon ^2-1\right) e^{2 d_1 |k_x|}+(\varepsilon -1)^2 \left(-e^{2 d_3 |k_x|}\right)+(\varepsilon +1)^2}{|k_x| \left((\varepsilon -1)^2 \left(e^{|k_x| (d_1+d_2)}-e^{d_3 |k_x|}\right)^2-(\varepsilon +1)^2 \left(e^{d |k_x|}-1\right)^2\right)} (1-r) a_a  \\
        &+ \frac{2 e^{\frac{1}{2} |k_x| (d_1-d_2+d_3)} \left((\varepsilon +1) e^{|k_x| (d_1+2 d_2+d_3)}-(\varepsilon -1) e^{|k_x| (2 d_1+d_2)}+(\varepsilon -1) e^{|k_x| (d_1+d_3)}-(\varepsilon +1) e^{d_2 |k_x|}\right)}{|k_x| \left((\varepsilon -1)^2 \left(e^{|k_x| (d_1+d_2)}-e^{d_3 |k_x|}\right)^2-(\varepsilon +1)^2 \left(e^{d |k_x|}-1\right)^2\right)} \sqrt{\varepsilon}t a_a \\
        b_{a-} =& \frac{-4 \varepsilon  e^{|k_x| (d_1+d_2+d_3)}-\left(\varepsilon ^2-1\right) e^{2 |k_x| (d_2+d_3)}+\left(\varepsilon ^2-1\right) e^{2 d_2 |k_x|}+(\varepsilon -1)^2 \left(-e^{2 d_3 |k_x|}\right)+(\varepsilon +1)^2}{|k_x| \left((\varepsilon -1)^2 \left(e^{|k_x| (d_1+d_2)}-e^{d_3 |k_x|}\right)^2-(\varepsilon +1)^2 \left(e^{d |k_x|}-1\right)^2\right)} (1-r) a_a \\
        & + \frac{2 e^{-\frac{1}{2} |k_x| (d_1-3 (d_2+d_3))} \left((\varepsilon -1) \left(-e^{|k_x| (d_1+d_2-d_3)}\right)+(\varepsilon +1) \left(e^{2 d_1 |k_x|}-e^{|k_x| (d_1-d_2-d_3)}\right)+\varepsilon -1\right)}{|k_x| \left((\varepsilon -1)^2 \left(e^{|k_x| (d_1+d_2)}-e^{d_3 |k_x|}\right)^2-(\varepsilon +1)^2 \left(e^{d |k_x|}-1\right)^2\right)} \sqrt{\varepsilon}t a_a
\end{split}
\end{equation}
}
 
{\tiny
\begin{equation}
\begin{split}        
        c_{a+} =& \frac{2 \varepsilon  e^{-2 d_2 |k_x|} \left(-(\varepsilon -1) e^{|k_x| (d_1+d_2+d_3)}-(\varepsilon +1) e^{|k_x| (d_1+3 d_2+d_3)}+(\varepsilon -1) e^{2 |k_x| (d_1+d_2)}+(\varepsilon +1) e^{2 d_2 |k_x|}\right)}{|k_x| \left((\varepsilon -1)^2 \left(e^{|k_x| (d_1+d_2)}-e^{d_3 |k_x|}\right)^2-(\varepsilon +1)^2 \left(e^{d |k_x|}-1\right)^2\right)} (1-r) a_a \\
        & + \frac{e^{-\frac{1}{2} |k_x| (d_1+3 d_2+d_3)} \left(-\left(\varepsilon ^2-1\right) e^{|k_x| (3 d_1+2 d_2+d_3)}+4 \varepsilon  e^{|k_x| (d_1+2 d_2+d_3)}+(\varepsilon -1)^2 e^{2 d_1 |k_x|+3 d_2 |k_x|}+\left(\varepsilon ^2-1\right) e^{|k_x| (d_1+d_3)}-(\varepsilon +1)^2 e^{d_2 |k_x|}\right)}{|k_x| \left((\varepsilon -1)^2 \left(e^{|k_x| (d_1+d_2)}-e^{d_3 |k_x|}\right)^2-(\varepsilon +1)^2 \left(e^{d |k_x|}-1\right)^2\right)} \sqrt{\varepsilon}t a_a \\
        c_{a-} =& \frac{2 \varepsilon  e^{|k_x| (d_2+d_3)} \left(-(\varepsilon +1) e^{|k_x| (2 d_1+d_2+d_3)}+(\varepsilon -1) e^{|k_x| (d_1+2 d_2)}+(\varepsilon +1) e^{d_1 |k_x|}-(\varepsilon -1) e^{|k_x| (d_2+d_3)}\right)}{|k_x| \left((\varepsilon -1)^2 \left(e^{|k_x| (d_1+d_2)}-e^{d_3 |k_x|}\right)^2-(\varepsilon +1)^2 \left(e^{d |k_x|}-1\right)^2\right)}(1-r) a_a \\
        & + \frac{e^{-\frac{1}{2} |k_x| (d_1-3 (d_2+d_3))} \left((\varepsilon -1)^2 e^{|k_x| (3 d_1+d_2-d_3)}-(\varepsilon +1)^2 e^{|k_x| (d_1-d_2-d_3)}-\left(\varepsilon ^2-1\right) e^{2 |k_x| (d_1+d_2)}+4 \varepsilon  e^{2 d_1 |k_x|}+\varepsilon ^2-1\right)}{|k_x| \left((\varepsilon -1)^2 \left(e^{|k_x| (d_1+d_2)}-e^{d_3 |k_x|}\right)^2-(\varepsilon +1)^2 \left(e^{d |k_x|}-1\right)^2\right)} \sqrt{\varepsilon}t a_a
\end{split}
\end{equation}
}
 
{\tiny
\begin{equation}
\begin{split}
        \Gamma_{a+} =& \frac{-4 \varepsilon  e^{\sqrt{k_{px}^2} (d_1+d_2+d_3)}-\left(\varepsilon ^2-1\right) e^{2 \sqrt{k_{px}^2} (d_1+d_3)}+\left(\varepsilon ^2-1\right) e^{2 d_1 \sqrt{k_{px}^2}}+(\varepsilon -1)^2 \left(-e^{2 d_3 \sqrt{k_{px}^2}}\right)+(\varepsilon +1)^2}{k_{px} \left(2 (\varepsilon -1)^2 \left(e^{\sqrt{k_{px}^2} (d_1+d_2)}-e^{ \sqrt{k_{px}^2}d_3} \right) \left((d_1+d_2) e^{\sqrt{k_{px}^2} (d_1 + d_2)} - d_3 e^{ \sqrt{k_{px}^2} d_3}\right)-2  (\varepsilon +1)^2 \left(e^{\sqrt{k_{px}^2} d}-1\right) d e^{\sqrt{k_{px}^2}d}\right)} (1-r)  \\
        &+ \frac{2 e^{\frac{1}{2} \sqrt{k_{px}^2} (d_1-d_2+d_3)} \left((\varepsilon +1) e^{\sqrt{k_{px}^2} (d_1+2 d_2+d_3)}-(\varepsilon -1) e^{\sqrt{k_{px}^2} (2 d_1+d_2)}+(\varepsilon -1) e^{\sqrt{k_{px}^2} (d_1+d_3)}-(\varepsilon +1) e^{d_2 \sqrt{k_{px}^2}}\right)}{k_{px} \left(2 (\varepsilon -1)^2 \left(e^{\sqrt{k_{px}^2} (d_1+d_2)}-e^{ \sqrt{k_{px}^2}d_3} \right) \left((d_1+d_2) e^{\sqrt{k_{px}^2} (d_1 + d_2)} - d_3 e^{ \sqrt{k_{px}^2} d_3}\right)-2  (\varepsilon +1)^2 \left(e^{\sqrt{k_{px}^2} d}-1\right) d e^{\sqrt{k_{px}^2}d}\right)} \sqrt{\varepsilon}t \\
        \Gamma_{a-} =& \frac{-4 \varepsilon  e^{\sqrt{k_{px}^2} (d_1+d_2+d_3)}-\left(\varepsilon ^2-1\right) e^{2 \sqrt{k_{px}^2} (d_2+d_3)}+\left(\varepsilon ^2-1\right) e^{2 d_2 \sqrt{k_{px}^2}}+(\varepsilon -1)^2 \left(-e^{2 d_3 \sqrt{k_{px}^2}}\right)+(\varepsilon +1)^2}{k_{px} \left(2 (\varepsilon -1)^2 \left(e^{\sqrt{k_{px}^2} (d_1+d_2)}-e^{ \sqrt{k_{px}^2}d_3} \right) \left((d_1+d_2) e^{\sqrt{k_{px}^2} (d_1 + d_2)} - d_3 e^{ \sqrt{k_{px}^2} d_3}\right)-2  (\varepsilon +1)^2 \left(e^{\sqrt{k_{px}^2} d}-1\right) d e^{\sqrt{k_{px}^2}d}\right)} (1-r) \\
        & + \frac{2 e^{-\frac{1}{2} \sqrt{k_{px}^2} (d_1-3 (d_2+d_3))} \left((\varepsilon -1) \left(-e^{\sqrt{k_{px}^2} (d_1+d_2-d_3)}\right)+(\varepsilon +1) \left(e^{2 d_1 \sqrt{k_{px}^2}}-e^{\sqrt{k_{px}^2} (d_1-d_2-d_3)}\right)+\varepsilon -1\right)}{k_{px} \left(2 (\varepsilon -1)^2 \left(e^{\sqrt{k_{px}^2} (d_1+d_2)}-e^{ \sqrt{k_{px}^2}d_3} \right) \left((d_1+d_2) e^{\sqrt{k_{px}^2} (d_1 + d_2)} - d_3 e^{ \sqrt{k_{px}^2} d_3}\right)-2  (\varepsilon +1)^2 \left(e^{\sqrt{k_{px}^2} d}-1\right) d e^{\sqrt{k_{px}^2}d}\right)} \sqrt{\varepsilon}t
\end{split}
\end{equation}
}   
 
{\tiny
\begin{equation}
\begin{split}        
        \Lambda_{a+} =& \frac{2 \varepsilon  e^{-2 d_2 \sqrt{k_{px}^2}} \left(-(\varepsilon -1) e^{\sqrt{k_{px}^2} d}-(\varepsilon +1) e^{\sqrt{k_{px}^2} (d_1+3 d_2+d_3)}+(\varepsilon -1) e^{2 \sqrt{k_{px}^2} (d_1+d_2)}+(\varepsilon +1) e^{2 d_2 \sqrt{k_{px}^2}}\right)}{k_{px} \left(2 (\varepsilon -1)^2 \left(e^{\sqrt{k_{px}^2} (d_1+d_2)}-e^{ \sqrt{k_{px}^2}d_3} \right) \left((d_1+d_2) e^{\sqrt{k_{px}^2} (d_1 + d_2)} - d_3 e^{ \sqrt{k_{px}^2} d_3}\right)-2  (\varepsilon +1)^2 \left(e^{\sqrt{k_{px}^2} d}-1\right) d e^{\sqrt{k_{px}^2}d}\right)} (1-r) \\
        & + \frac{e^{-\frac{1}{2} \sqrt{k_{px}^2} (d + 2 d_2)} \left(-\left(\varepsilon ^2-1\right) e^{\sqrt{k_{px}^2} (3 d_1+2 d_2+d_3)}+4 \varepsilon  e^{\sqrt{k_{px}^2} (d_1+2 d_2+d_3)}+ (\varepsilon -1)^2 e^{\sqrt{k_{px}^2}(2 d_1 +3 d_2)} + \left(\varepsilon ^2-1\right) e^{\sqrt{k_{px}^2} (d_1+d_3)}-(\varepsilon +1)^2 e^{ \sqrt{k_{px}^2}d_2}\right)}{k_{px} \left(2 (\varepsilon -1)^2 \left(e^{\sqrt{k_{px}^2} (d_1+d_2)}-e^{ \sqrt{k_{px}^2}d_3} \right) \left((d_1+d_2) e^{\sqrt{k_{px}^2} (d_1 + d_2)} - d_3 e^{ \sqrt{k_{px}^2} d_3}\right)-2  (\varepsilon +1)^2 \left(e^{\sqrt{k_{px}^2} d}-1\right) d e^{\sqrt{k_{px}^2}d}\right)} \sqrt{\varepsilon}t \\
        \Lambda_{a-} =& \frac{2 \varepsilon  e^{\sqrt{k_{px}^2} (d_2+d_3)} \left(-(\varepsilon +1) e^{\sqrt{k_{px}^2} (2 d_1+d_2+d_3)}+(\varepsilon -1) e^{\sqrt{k_{px}^2} (d_1+2 d_2)}+(\varepsilon +1) e^{d_1 \sqrt{k_{px}^2}}-(\varepsilon -1) e^{\sqrt{k_{px}^2} (d_2+d_3)}\right)}{k_{px} \left(2 (\varepsilon -1)^2 \left(e^{\sqrt{k_{px}^2} (d_1+d_2)}-e^{ \sqrt{k_{px}^2}d_3} \right) \left((d_1+d_2) e^{\sqrt{k_{px}^2} (d_1 + d_2)} - d_3 e^{ \sqrt{k_{px}^2} d_3}\right)-2  (\varepsilon +1)^2 \left(e^{\sqrt{k_{px}^2} d}-1\right) d e^{\sqrt{k_{px}^2}d}\right)}(1-r) \\
        & + \frac{e^{-\frac{1}{2} \sqrt{k_{px}^2} (d_1-3 (d_2+d_3))} \left((\varepsilon -1)^2 e^{\sqrt{k_{px}^2} (3 d_1+d_2-d_3)}-(\varepsilon +1)^2 e^{\sqrt{k_{px}^2} (d_1-d_2-d_3)}-\left(\varepsilon ^2-1\right) e^{2 \sqrt{k_{px}^2} (d_1+d_2)}+4 \varepsilon  e^{2 d_1 \sqrt{k_{px}^2}}+\varepsilon ^2-1\right)}{k_{px} \left(2 (\varepsilon -1)^2 \left(e^{\sqrt{k_{px}^2} (d_1+d_2)}-e^{ \sqrt{k_{px}^2}d_3} \right) \left((d_1+d_2) e^{\sqrt{k_{px}^2} (d_1 + d_2)} - d_3 e^{ \sqrt{k_{px}^2} d_3}\right)-2  (\varepsilon +1)^2 \left(e^{\sqrt{k_{px}^2} d}-1\right) d e^{\sqrt{k_{px}^2}d}\right)} \sqrt{\varepsilon}t
\end{split}
\end{equation}
}
 
{\tiny
\begin{equation}
\begin{split}
        \Lambda_{a+}^{'} =& \frac{2 \varepsilon  e^{-2 d_2 \sqrt{k_{px}^2}} \left(-(\varepsilon -1) e^{\sqrt{k_{px}^2} d}-(\varepsilon +1) e^{\sqrt{k_{px}^2} (d_1+3 d_2+d_3)}+(\varepsilon -1) e^{2 \sqrt{k_{px}^2} (d_1+d_2)}+(\varepsilon +1) e^{2 d_2 \sqrt{k_{px}^2}}\right)}{k_{px} \left(2 (\varepsilon -1)^2 \left(e^{\sqrt{k_{px}^2} (d_1+d_2)}-e^{ \sqrt{k_{px}^2}d_3} \right) \left((d_1+d_2) e^{\sqrt{k_{px}^2} (d_1 + d_2)} - d_3 e^{ \sqrt{k_{px}^2} d_3}\right)-2  (\varepsilon +1)^2 \left(e^{\sqrt{k_{px}^2} d}-1\right) d e^{\sqrt{k_{px}^2}d}\right)} \\
        & + \frac{e^{-\frac{1}{2} \sqrt{k_{px}^2} (d + 2 d_2)} \left(-\left(\varepsilon ^2-1\right) e^{\sqrt{k_{px}^2} (3 d_1+2 d_2+d_3)}+4 \varepsilon  e^{\sqrt{k_{px}^2} (d_1+2 d_2+d_3)}+ (\varepsilon -1)^2 e^{\sqrt{k_{px}^2}(2 d_1 +3 d_2)} + \left(\varepsilon ^2-1\right) e^{\sqrt{k_{px}^2} (d_1+d_3)}-(\varepsilon +1)^2 e^{ \sqrt{k_{px}^2}d_2}\right)}{k_{px} \left(2 (\varepsilon -1)^2 \left(e^{\sqrt{k_{px}^2} (d_1+d_2)}-e^{ \sqrt{k_{px}^2}d_3} \right) \left((d_1+d_2) e^{\sqrt{k_{px}^2} (d_1 + d_2)} - d_3 e^{ \sqrt{k_{px}^2} d_3}\right)-2  (\varepsilon +1)^2 \left(e^{\sqrt{k_{px}^2} d}-1\right) d e^{\sqrt{k_{px}^2}d}\right)} \varepsilon \\
        \Lambda_{a-}^{'} =& \frac{2 \varepsilon  e^{\sqrt{k_{px}^2} (d_2+d_3)} \left(-(\varepsilon +1) e^{\sqrt{k_{px}^2} (2 d_1+d_2+d_3)}+(\varepsilon -1) e^{\sqrt{k_{px}^2} (d_1+2 d_2)}+(\varepsilon +1) e^{d_1 \sqrt{k_{px}^2}}-(\varepsilon -1) e^{\sqrt{k_{px}^2} (d_2+d_3)}\right)}{k_{px} \left(2 (\varepsilon -1)^2 \left(e^{\sqrt{k_{px}^2} (d_1+d_2)}-e^{ \sqrt{k_{px}^2}d_3} \right) \left((d_1+d_2) e^{\sqrt{k_{px}^2} (d_1 + d_2)} - d_3 e^{ \sqrt{k_{px}^2} d_3}\right)-2  (\varepsilon +1)^2 \left(e^{\sqrt{k_{px}^2} d}-1\right) d e^{\sqrt{k_{px}^2}d}\right)}\\
        & + \frac{e^{-\frac{1}{2} \sqrt{k_{px}^2} (d_1-3 (d_2+d_3))} \left((\varepsilon -1)^2 e^{\sqrt{k_{px}^2} (3 d_1+d_2-d_3)}-(\varepsilon +1)^2 e^{\sqrt{k_{px}^2} (d_1-d_2-d_3)}-\left(\varepsilon ^2-1\right) e^{2 \sqrt{k_{px}^2} (d_1+d_2)}+4 \varepsilon  e^{2 d_1 \sqrt{k_{px}^2}}+\varepsilon ^2-1\right)}{k_{px} \left(2 (\varepsilon -1)^2 \left(e^{\sqrt{k_{px}^2} (d_1+d_2)}-e^{ \sqrt{k_{px}^2}d_3} \right) \left((d_1+d_2) e^{\sqrt{k_{px}^2} (d_1 + d_2)} - d_3 e^{ \sqrt{k_{px}^2} d_3}\right)-2  (\varepsilon +1)^2 \left(e^{\sqrt{k_{px}^2} d}-1\right) d e^{\sqrt{k_{px}^2}d}\right)} \varepsilon
\end{split}
\end{equation}
}

\bibliography{reference, plasmonic_gratings}
 
\end{document}